\documentclass{aa}  
\usepackage{graphicx}
\usepackage{txfonts}
\usepackage{longtable,ltcaption}
\usepackage{lscape}
\usepackage{natbib}
\usepackage{lscape}
\usepackage{amsmath}
\usepackage{wasysym}
\usepackage{graphics}
\usepackage{txfonts}
\usepackage{color}
\usepackage{times}
\usepackage{parskip}
\usepackage{pdflscape}
\usepackage{geometry}
\usepackage{marginnote}
\usepackage[table]{xcolor}
\usepackage{tabu}
\usepackage{stfloats}
\usepackage{multicol}
\usepackage{amsfonts}

\makeatother

\newfont{\nlx}{cmssdc10 scaled 900}
\newfont{\mfont}{cmssdc10 scaled 760}
\definecolor{myblue1}{rgb}{0.0,0.604,0.831} 
\definecolor{myblue2}{rgb}{0.0,0.49,0.6745}
\definecolor{myblue3}{rgb}{0.0156,0.4078,0.9921}
\definecolor{myblue4}{rgb}{0.0,0.44,0.87}
\definecolor{myred1}{rgb}{0.529,0.019,0.017}
\definecolor{mycyan}{rgb}{0.63921569,0.0,0.48235294}

\newcommand{\brem}[1]{\textcolor{black}{\nlx #1}}

\newcommand{\Starlight}{{\sc Starlight}}

\def\msun{$\mathrm{M}_{\odot}$}
\def\zsun{$\mathrm{Z}_{\odot}$}

\def\thalf{$t_{1/2}$}
\def\D4000{$D_{4000}$}

\def\rr{$R^{\star}$}
\def\rbulge{$R_{\rm B}$}
\newcommand{\sbb}{mag/$\sq\arcsec$}
\newcommand{\dmb}{$<\!\!\!\delta\mu_{9{\rm G}}\!\!\!>$}

\def\mgr{{\sc mgr}}
\def\mit{$\tau_{\rm m}$}
\def\RY{${\cal RY}$}
\def\mstar{${\cal M}_{\star}$}
\def\mbstar{${\cal M}_{\star,\textrm{B}}$}
\def\mstotal{${\cal M}_{\star,\textrm{T}}$}
\def\tstar{$t_{\star}$}

\def\tottmass{$\langle t_{\star} \rangle_{{\cal M}}$}
\def\tottlight{$\langle t_{\star} \rangle_{{\cal L}}$}
\def\totzmass{$\langle Z_{\star} \rangle_{{\cal M}}$}
\def\totzlight{$\langle Z_{\star} \rangle_{{\cal L}}$}
\def\mtmass{$\langle t_{\star,\textrm{B}} \rangle_{{\cal M}}$}
\def\utmass{$\langle t_{\star} \rangle_{{\cal M}}$}
\def\utlight{$\langle t_{\star} \rangle_{{\cal L}}$}

\def\zgas{$Z_{\rm g}$}
\def\zstar{$Z_{\star}$}

\def\mzmass{$\langle Z_{\star,\textrm{B}} \rangle_{{\cal M}}$}

\def\tsstar{$\Sigma_{\star}$}
\def\sstar{$\Sigma_{\star,\mathrm{B}}$}
\newcommand{\PutLabel}[3]{\put(#1,#2){#3}}
\newfont{\hvss}{cmssdc10 scaled 1540}
\def\?{{\bf\color{red}?}}
\def\mdtmass{$\langle t_{\star,\textrm{D}} \rangle_{{\cal M}}$}
\def\mdzmass{$\langle Z_{\star,\textrm{D}} \rangle_{{\cal M}}$}
\def\reff{$R_{\rm eff}$}
\def\ha{H$\alpha$}
\def\hb{H$\beta$}

\def\dbdt{$\delta t_{\rm BD}$}
\def\dbdz{$\delta Z_{\rm BD}$}
\newcommand{\PutWin}[4]{
\put(#1,#2){\parbox{#3}{#4}}}

\begin{document} 

\title{The continuous rise of bulges out of galactic disks}
   \author{Iris Breda
          \inst{1,2}
          \and
          Polychronis Papaderos
          \inst{1}
          }
   \institute{Instituto de Astrof\'{i}sica e Ci\^{e}ncias do Espaço - Centro de Astrof\'isica da Universidade do Porto, Rua das Estrelas, 4150-762 Porto, Portugal\\
         \and
             Departamento de F\'isica e Astronomia, Faculdade de Ci\^encias, Universidade do Porto, Rua do Campo Alegre, 4169-007 Porto, Portugal\\
             \email{iris.breda@astro.up.pt, papaderos@astro.up.pt}
             }

   \date{Received 03 August 2017; accepted 30 November 2017}

\abstract
{A key subject in extragalactic astronomy concerns the chronology and driving mechanisms of bulge formation in late-type galaxies (LTGs). The standard scenario distinguishes between classical bulges and pseudo-bulges (CBs and PBs, respectively), the first thought to form monolithically prior to disks and the second gradually out of disks. These two bulge formation routes obviously yield antipodal predictions on the bulge age and bulge-to-disk age contrast, both expected to be high (low) in CBs (PBs).}
{Our main goal is to explore whether bulges in present-day LTGs segregate into two evolutionary distinct classes, as expected from the standard scenario. Other questions motivating this study center on evolutionary relations between LTG bulges and their hosting disks, and the occurrence of accretion-powered nuclear activity as a function of bulge stellar mass \mstar\ and stellar surface density \tsstar.}
{In this study we have combined three techniques -- surface photometry, spectral modeling of integral field spectroscopy data and suppression of stellar populations younger than an adjustable age cutoff with the code {\sc RemoveYoung} (\RY) -- toward  a systematic analysis of the physical and evolutionary properties (e.g., \mstar, \tsstar\ and mass-weighted stellar age \utmass\ and metallicity \totzmass, respectively) 
of a representative sample of 135 nearby ($\leq$130 Mpc) LTGs from the CALIFA survey that cover a range between 10$^{8.9}$ \msun\ and 10$^{11.5}$ \msun\ in total stellar mass \mstotal.
In particular, the analysis here revolves around \dmb, a new distance- and formally extinction-independent measure of the contribution by 
stellar populations of age $\geq$9~Gyr to the mean $r$-band surface brightness of the bulge. 
We argue that \dmb\ offers a handy semi-empirical tracer of the physical and evolutionary properties
of LTG bulges and a promising means for their characterization.
}
{The essential insight from this study is that LTG bulges form over three dex in \mstar\ and more than one dex in \tsstar\ 
a tight continuous sequence of increasing \dmb\ with increasing \mstar, \tsstar, \utmass\ and \totzmass. 
Along this continuum of physical and evolutionary properties, our sample spans a range of $\sim$4~mag in \dmb: 
high-\dmb\ bulges are the oldest, densest and most massive ones (\utmass$\sim$11.7 Gyr, \tsstar $>10^9$ \msun\,kpc$^{-2}$,
\mstar $\geq 10^{10}$ \msun), whereas the opposite is the case for low-\dmb\ bulges (\utmass$\sim$7 Gyr) that generally reside in low-mass LTGs.
Furthermore, we find that the bulge-to-disk age and metallicity contrast, as well as the bulge-to-disk mass ratio, show a positive trend 
with \mstotal, raising from, respectively, $\sim$0~Gyr, $\sim$0~dex and 0.25 to $\sim$3 Gyr, $\sim$0.3~dex and 0.67 
across the mass range covered by our sample.
Whereas gas excitation in lower-mass ($\lesssim 10^{9.7}$ \msun) bulges is invariably dominated by star formation (SF), 
LINER- and Seyfert-specific emission-line ratios were exclusively documented in high-mass ($\gtrsim 10^{10}$ \msun), 
high-\tsstar\ ($\gtrsim 10^{9}$ \msun\ kpc$^{-2}$) bulges. This is in agreement with previous work and 
consistent with the notion that the Eddington ratio or the black~hole-to-bulge mass ratio scale with \mstar.
The coexistence of Seyfert and SF activity in $\sim$20\% of higher-\mstar, high-\tsstar\ bulges being spectroscopically classified 
as Composites suggests that the onset of AGN-driven feedback does not necessarily lead to an abrupt termination of SF in LTG nuclei.
}
{The continuity both in the properties of LTG bulges themselves and in their age and metallicity contrast to their parent disks suggests that these components evolve alongside in a concurrent process that leads to a continuum of physical and evolutionary characteristics. Our results are consistent with a picture where bulge growth in LTGs is driven by a superposition of quick-early and slow-secular processes, the relative importance of which increases with \mstotal.
These processes, which presumably combine in situ SF in the bulge and inward migration of material from the disk,
are expected to lead to a non-homologous radial growth of \tsstar\ and a trend for an increasing S\'{e}rsic index with increasing galaxy mass.
}

\maketitle

\parskip = \baselineskip

\section{Introduction \label{intro}}
The driving mechanisms and chronology of the buildup of bulges in late-type galaxies (LTGs) 
is an issue of key relevance to our understanding of galaxy evolution.
According to our current knowledge on bulge demographics in the local universe, a large fraction of LTGs host pseudo-bulges \citep[PBs; e.g.,][]{Gad09,FisDro11,FerLor14} that substantially differ from classical bulges (CBs) in their spectrophotometric and kinematical characteristics. The latter resemble in many respects 'old and dead' elliptical galaxies, lacking ongoing star-formation (SF), exhibit a spheroidal shape with inwardly steeply increasing surface brightness profiles (SBPs) being well approximated by the \citet{Sersic63} fitting law with a high ($\gtrsim 3$) exponent $\eta$, show stellar kinematics dominated by velocity dispersion ($\sigma_{\star}$) and obey the \citet{Kormendy77} scaling relations 
for normal elliptical galaxies \citep{FisDro10}. 
It is observationally established that CBs contain a super-massive black hole (SMBH) with a mass $M_{\bullet}$ tightly correlating 
with their stellar mass \mbstar, $\sigma_{*}$ and optical luminosity \citep[][see also, Ferrarese \& Merritt 2000]{Ho2008,KormendyHo2013}. 
Traditionally, bulges were thought to invariably form early-on via violent quasi-monolithic gas collapse \citep{Lar74} or mergers \citep{BBF92,Agu01,KelNus12} associated with vigorous nuclear starbursts \citep{Oka12}, with the disk gradually building up around them. Whereas this inside-out galaxy formation scenario appears consistent with important integral characteristics of CBs (e.g., their red colors), it does not offer a plausible explanation for the presence of PBs in present-day LTGs. 
These generally show ongoing SF, a significant degree of rotational support \citep[][for a review]{KorKen04} and flatter/ellipsoidal shapes with nearly exponential SBPs \citep[$\eta \la 2$; e.g.,][]{DroFis07,FisDro10}. Even though there is observational evidence that PBs also contain a SMBH \citep{Kor11,KormendyHo2013}, in some cases revealing itself as an active galactic nucleus \citep[AGN; e.g.,][see Kormendy \& Ho 2013 for a review]{Kotilainen16}, these do not follow the $M_{\bullet}$-$\sigma_{*}$ correlation for CBs, which appears to be
consistent with a different formation route. Indeed, the prevailing concept on PB formation is that these entities emerge gradually out of galactic disks through gentle gas inflow spawning quasi-continuous SF and the emergence of a central bulge-like luminosity excess at their centers \citep[e.g.,][]{CdJB96,Car01,KorKen04}. Besides bar-driven gas inflow \citep[e.g.,][]{SprHer05}, various other mechanisms, such as inward stellar migration, minor mergers with low-mass satellites, or a purely dynamical re-arrangement of the disk \citep{Scannapieco10,Guedes13,Bird12,Roskar12,Grand14,Halle15} have been proposed as further contributors to PB growth along the Gyr-long secular evolution of LTGs.

These two scenarios, the first one envisaging bulge formation prior to disks and the second one out of disks obviously yield antipodal predictions on the bulge age and bulge-to-disk color contrast, both expected to be high (low) in CBs (PBs). Naively, one might therefore expect bulges in present-day LTGs to describe a bimodal age distribution, with each of the two classes showing a high degree of homogeneity in its spectrophotometric, chemodynamical and evolutionary properties, echoing two distinct formation routes. However, the available observational data contrast the picture of an age bimodality in LTG bulges, suggesting instead a substantial spread, if not smooth transition, in their properties across their mass and luminosity range. For example, \cite{Wys97}, reviewing the subject two decades ago, conclude that bulges show a considerable heterogeneity, with merely higher luminosity ones (CBs) having a closer affinity with Ellipticals and lower luminosity bulges (PBs) with disks.

From the perspective of the monolithic bulge formation scenario, there is evidently no other option than to interpret PBs as SF-rejuvenated CBs, and indeed, this proposal has been put forward in several studies \citep[e.g.,][]{ThoDav06,Johnston12,Johnston14,Mor12,Mor16}. 
For example, \cite{ThoDav06} drew this conclusion by reproducing a subset of Lick indices 
(H$\beta$, H$\gamma\,{\rm A}$, H$\delta_{\rm A}$, [MgFe]\arcmin, and <Fe>) 
for local bulges with a two-component evolutionary synthesis (ESS) model that involves a 15 Gyr old stellar component, 
which 1-2 Gyr ago underwent a SF episode producing up to 10-30\% of its total \mstar. On the basis, of these models and 
observed correlations between central $\sigma_{\star}$ (aka \mstotal) and the luminosity-weighted age, metallicity and $\alpha$/Fe ratio, 
these authors argue that the smallest bulges are the youngest and have experienced a late iron enrichment by type~Ia SNe. 

Intermediate between the 'quick' (monolithic) and 'slow' (secular) bulge formation scenarios above is a set of models envisaging the dominant phase of bulge formation to occur 
in a prolonged episode of $\sim$0.3-0.8 Gyr through inward migration and coalescence of massive ($\ga 10^{8-9}$ \msun) SF clumps forming continuously in the disk out of gas  instabilities \citep{Noguchi99,Bournaud07,Car07,Elm08,Mandelker14,Mandelker17}. Observational support for this picture comes from the clumpiness of high-$z$ proto-disks and the estimated \mstar\ of their SF clumps \citep{FS11,Wuyts12}. 
Additionally, inflowing inter-clump gas \citep[e.g.,][]{Hopkins2012,Zolotov15} could sustain perpetual rejuvenation of the bulge with in situ SF, which acts together with stellar migration, minor mergers with dwarf satellites and other dynamical effects \citep[e.g.,][]{Bird12,Roskar12,Guedes13,Grand14,Halle15} toward bulge buildup in the ensuing Gyr of galactic evolution. Depending on the timescales and the contribution of these different processes, it might be expected this set of prolonged bulge formation scenarios to yield a range of stellar ages in present-day LTG bulges.  

From the observational point of view, the notion of a correlated evolution of disk and bulge has been put forward is several studies, mostly on the basis of photometric investigations of local LTGs \citep[e.g.,][]{GA01,Gad09}. 
For example, early work by \citet{PelBal96} finds that "color variations from galaxy to galaxy are much larger than color differences between disk and bulge in each galaxy", following that "the underlying old population of disks and bulges is much more similar" than previously thought. 
Also, other works find the color contrast between bulge and disk to be relatively small \citep[e.g., 0.3 $g$-$i$ mag for isolated galaxies studied in][]{FerLor14} and interpret this as evidence for a correlated evolution of these components \citep[see also, e.g.,][]{CdJB96,GA01}.
Similarly, a photometric study by \citet{Car07} finds that the colors of bulges are correlated with those of the disks in which they are embedded, suggesting that an early phase in bulge formation must have been supplemented through "continuing" rejuvenation, and in some cases even that the bulk of the stellar mass in bulges has assembled more recently than the disk. These authors also point out that the scaling relations between bulge stellar age and bulge/galaxy mass hint at similar formation processes for all components, suggesting that bulges across their entire range in mass and age result from the internal evolution of the parent disks. They also show that dynamical friction of massive clumps in gas-rich disks offers an explanation for the formation of late-type bulges, especially for those that are older than their surrounding disks, consistently with the picture above. Coming to our own Milky Way (MW), \cite{Nes14} demonstrate by means of N-body \& smooth particle hydrodynamics simulations that the presence of young stars predominantly near the plane is expected for a bulge that has formed from the disk via dynamical instabilities, whereas it cannot be accounted for by monolithic collapse.

The picture of a concomitant evolution of bulge and disk has also received support from HST imaging studies of MW analogs at higher $z$: \citet{vD13}, by applying the abundance matching technique to MW-progenitor candidates out to $z$ = 2.5 point out that there is an absence of high-density "naked bulges" at $z\sim 2$, around which disks subsequently assemble \citep[see also][]{Patel13}. 
These authors also find that MW-like LTGs have built $\sim$90\% of their present \mstar\ since $z$=2.5 with most of the SF occurring before $z$=1. 
They verify that for $1 < z < 2.5$ the mass in the central 2 kpc of MW progenitors increases by a factor of $3.2^{+0.8}_{-0.7}$, which rules out models in which bulges were fully assembled first and disks gradually formed around them. 
In this context, the question of how the bulge and the disk grew relative to each other since this early cosmic epoch is of considerable interest. \citet{Mar16} find from HST photometry of 1495 massive galaxies in the CANDELS field that two-component (bulge \& disk) systems in the redshift interval $1.5 < z < 3$ roughly maintain their bulge size whereas their disks grow by a factor
of approximately 3. A subsequent study of such two-component galaxies in \citet{Mar17} finds that the \mstar\ enclosed within the bulge and disk remains nearly constant over the latter redshift interval. This, and the fact that the disk shows a higher star formation rate (SFR) than the bulge led these authors to conclude that about one half of the stellar mass formed in the disk must have migrated into the bulge, promoting its gradual growth. We note though that evolution of the bulge-to-disk mass ratio across $z$ remains a subject of investigation. For example. \cite{Tac17} report that galaxies with \mstar$\leq 10^{11}$ \msun\ have since $z\sim$2 on average doubled their stellar mass throughout their radial extent.

Whereas most lines of evidence point to an interwoven evolution of bulge and disk, our understanding on how this process has shaped the heterogeneity of present-day bulges is far from complete. In fact, neither observations nor theory yield as yet clear-cut discriminators between CBs and PBs, or even unambiguous evidence for an evolutionary dichotomy between both.
From the photometric point of view, it is common practice to classify bulges by their S\'ersic index $\eta$, which is assumed to be $>$ ($\leq$) $\sim$2 for CBs (PBs). 
However, this cutoff has no clear physical foundation \citep[e.g.,][]{Gadotti12} and essentially rests on subjective considerations. 
For instance, \citet[][cf. their Fig. 7]{FisDro10} find that across the bulge sequence there is a substantial overlap between CBs and PBs, depending on whether they are selected according to mid-infrared colors or $\eta$. Likewise, some photometric classification clues for PBs merely rely on the absence of correlations that are known 
to apply to CBs: \citet{FisDro10} find that the $\eta$ of PBs is uncorrelated with other bulge structural properties, unlike CBs. This is in agreement with the analysis of \citet{Gad09} who concludes that CBs follow a correlation between $\eta$ and bulge-to-total (B/T) ratio, whereas PBs do not, and they actually occupy on the fundamental plane the same locus as disks.
In this regard, a proposal made by \citet{Gad09} is that a better separation between CBs and PBs is possible on the basis of deviations from the \citet{Kormendy77} relation
rather than on $\eta$. On the other hand, this study has shown that CBs are offset from Ellipticals in the mass-size relation, suggesting that the former are not simply Ellipticals surrounded by disks. The search for discriminators between CBs and PBs continues attracting considerable interest, with new schemes proposed, such as, for example, that by \citet{Neumann17}, which uses a combination of $\eta$, light concentration index $C_{20,50}$, the \citet{Kormendy77} relation and the inner slope of the radial $\sigma_{\star}$ profile.

As for spectral modeling studies of single-fiber SDSS data, they find a trend for increasing bulge age and metallicity with increasing
\mstotal, with CBs (PBs) generally populating the high (low) range of a broad sequence in mass \citep[e.g.,][]{Zhao12,Ribeiro16}.
These trends are essentially echoing relations obtained or corroborated over the past decade from spectral fitting of SDSS data for large extragalactic probes, as for example, a positive correlation between \mstotal, gas-phase and stellar metallicity (\zgas\ and \zstar, respectively) and light- and 
mass-weighted stellar age (\tottlight\ and \tottmass, respectively) \citep[e.g.,][]{Kauffmann13a,Tre04}.
They are also consistent with the galaxy 'downsizing' picture, reflected in, for example, an anti-correlation between \mstotal\ with specific star formation rate (sSFR) and \tottlight\ for local galaxies \citep[e.g.,][]{Bri04,Heavens04,Gal05,Noeske07a,Asari07}, which implies that massive galaxies have experienced the dominant phase of their assembly early on, whereas lower-mass systems build up their stellar mass over longer timescales. 
For instance, \citet{Gal05} find a sequence of increasing \mstotal\ and \tottlight\ with increasing light-weighted 
\zstar\ (\totzlight) and 4000 $\AA$ break strength, indicating that low-mass galaxies are typically younger and less metal-enriched, 
contrary to massive ones, with a transition between these two regimes occurring at $3\cdot 10^9 <$ \mstotal\ (\msun) $< 3\cdot 10^{10}$.
It is unclear though whether the absence of sensitive discriminators between CBs and PBs, despite intense exploration of spectroscopic SDSS data over the past years, is due to a genuine continuity in the physical and evolutionary properties (e.g., \mstar\ and \tottmass, respectively) of LTG bulges, or party because of aperture effects \citep[cf. discussion in, e.g.,][and references therein]{Gomes16b}. The latter are unavoidable, given that the 3\arcsec\ fiber of SDSS captures only a small portion of the bulge for nearby LTGs, whereas it also includes the surrounding disk for more distant ones. In the first case, evolutionary characteristics inferred from spectral modeling are representative for the bulge only as long as age and metallicity gradients therein are weak, whereas in the second case they could be systematically biased through the star-forming disk.
For the typical bulge radius of a massive LTG in our sample ($\sim$3 kpc; cf. Table~\ref{tab1}) this bias is expected for $z \geq 0.34$ (standard cosmology assumed).

Wide-field integral field spectroscopy (IFS) can in principle overcome these aperture biases, since it allows for spectroscopic analysis of the total bulge emission within a photometrically defined radius obtained from SBP decomposition into bulge and disk.
Although recent IFS studies have explored various physical relations between, for example, \tsstar, \tottlight, \totzlight\ and SFR for LTGs as a function of galactocentric radius \citep[e.g.,][]{Gon14,Gon16,CD16,Zibetti17}, they generally did not incorporate a structural analysis that would have permitted extraction and spectral modeling of the total bulge emission within a uniformly defined radius. An exception to this is the work by \citet{SB14} where the bulge extent was estimated from image decomposition as the radius where the intensity of the bulge equals that of the disk, to which that study was mainly devoted. The normalization of radial profiles for various quantities inferred from these IFS studies to the galaxy effective radius is another possible issue, since this approach bears the risk of comparing determinations within the bulge of higher-B/T LTGs with those in the disk of lower-B/T LTGs (cf. Sect.~\ref{SpectralModeling}).

The goal of this study is to explore on the basis of a combined spectral modeling and surface photometry analysis of a representative sample of local LTGs 
the connection between physical and evolutionary properties of bulges (e.g., \mstar, \tsstar\ and \tottmass, respectively) within a uniformly defined isophotal
radius that encompasses almost their total emission. 

It is motivated by the question of whether CBs and PBs are truly evolutionary distinct, or rather the opposite 
ends of a continuous sequence reflecting a decreasing relative contribution of 'quick and early' to 'slow and secular' processes to the bulge stellar mass growth.
If so, in which manner might the relative importance of these two processes be imprinted on the evolutionary and chemical properties of bulges in present-day LTGs, 
and can it be linked to a semi-empirical indicator that could ease bulge classification? 
Another goal of this study is to explore through standard emission-line diagnostics the connection between gas excitation mechanisms and evolutionary properties
of LTG bulges in order to gain insights into the occurrence of accretion-powered activity and its possible regulatory role on bulge growth.

To address these questions we extracted 135 LTGs from the CALIFA IFS survey \citep{Sanchez12-DR1,Sanchez16-DR3} which were analyzed using spectral synthesis models and SDSS surface photometry. In Sect. \ref{sampleandmethods} we describe the sample selection and the methodology employed for the spectral modeling and structural analysis of the galaxy sample. Section \ref{res1} provides an overview of the main results from this analysis, with a discussion following 
in Sect. \ref{dis}. The conclusions from this study are summarized in Sect. \ref{conc}.

\section{Data sample and methodology \label{sampleandmethods}}
\subsection{Sample selection \label{sample}}
\begin{figure}[htpb] 
\includegraphics[width=1\linewidth]{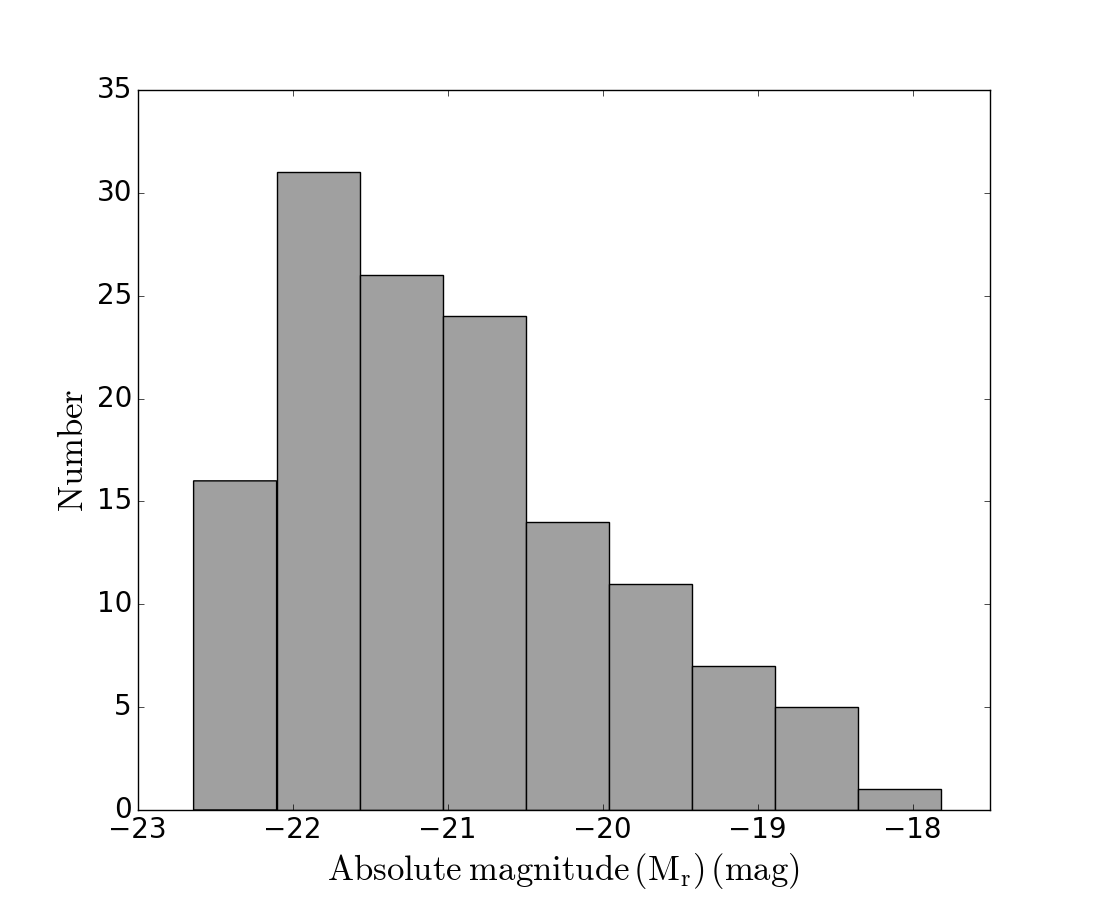}
\caption{Distribution of extinction-corrected total absolute magnitudes in the SDSS $r$ for the analyzed sample of 135 LTGs.} \label{histogram}
\end{figure}

The galaxy sample analyzed here was extracted from the 3$^{\rm rd}$ Data Release of the CALIFA integral field spectroscopy (IFS) survey \citep[667 galaxies;][see http://califa.caha.es]{Sanchez16-DR3}, and comprises 135 non-interacting, nearly face-on local ($\leq$130 Mpc) LTGs, selected for spatially resolved analysis of the physical and evolutionary properties of their bulge component. 
Systems strongly overlapping with bright foreground Galactic stars or extended background sources, or show signs of recent or ongoing interactions, or other morphological distortions (e.g., polar rings) were excluded. 
This is also the case for higher-inclination ($> 40^{\circ}$) LTGs in order to minimize internal obscuration effects and disk contamination. 
The sample may be considered representative for LTGs in the local universe, as it spans a range between 
--17.8 mag and --22.6 mag in SDSS \textit{r} absolute magnitude $M_r$ (Fig.~\ref{histogram}), as determined from 
integration of SBPs down to an extinction-corrected surface brightness level 24 \sbb, and covers all late-type morphological 
types.
With regard to the bulge component, the LTGs under study span a factor $\sim$250 in optical luminosity ($-20.6\la M_r\,\,\textrm{(mag)}\,\la -14.6$) 
and 3~dex in \mstar\ (10$^{8.3}$ -- 10$^{11.3}$ \msun; cf. Table~\ref{tab1} and Fig.~\ref{res}). 

\vspace*{-0.5cm}
\subsection{Data analysis \label{meth}}
The methodology adopted in this study combines spatially resolved modeling of IFS data with two further elements. The first one is the structural analysis of SDSS $r$ band images in order to obtain a uniformly defined and largely model-independent isophotal radius for the bulge, within which physical and evolutionary quantities from spectral synthesis were subsequently analyzed. This ensures a homogeneous extraction and spectral fitting of nearly the total bulge emission, free of aperture biases that are inherent to single-fiber spectroscopy. 
The second novel element of our study is the spaxel-by-spaxel post-processing of the spectral synthesis output with the code 
{\sc RemoveYoung} \citep[\RY;][]{GP16-RY} 
with the goal of a quantitative study of the contribution of stellar populations forming over the past 9 Gyr ($\sim$2/3 of the age of the Universe) to the optical surface brightness ($\mu$) and \tsstar\ of the bulge and disk. 
The insights gained from this computationally expensive task, which is performed for the first time on a large set of IFS data here, will be discussed in detail
in a forthcoming article.

This pilot analysis revolves around \dmb, a new distance-independent and, formally, also extinction-independent quantity obtained with \RY, 
which, as we discuss next, offers a semi-empirical tracer of the physical and evolutionary properties of LTG bulges and possibly 
a convenient means for their classification. 
\subsubsection{Determination of the bulge radius \label{SurfPhot}}
\begin{center}
\begin{figure}[htpb] 
\includegraphics[width=1\linewidth]{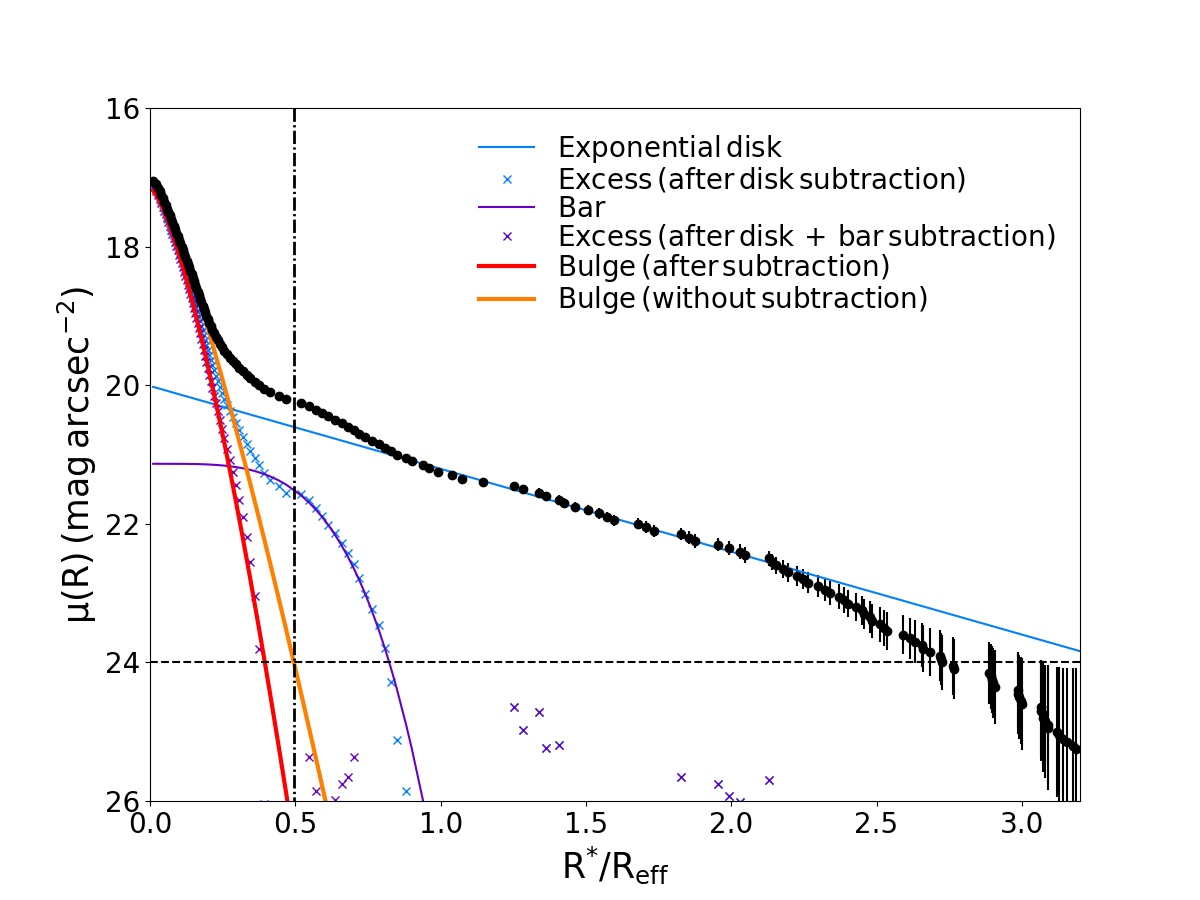}
\caption[Decomposition of the SBP of NGC 0776]{Illustrative example of the decomposition with iFit of the SDSS $r$-band SBP of the LTG \object{NGC 0776} 
(cf. Fig.~\ref{classes}) into bulge, bar and disk (red, magenta and blue, respectively). The bulge and bar are approximated by a S\'ersic model, 
and the down-bending (type~ii) disk by a pure exponential fitted to intermediate radii (1$\leq\mathrm{R^{\star}/R_{eff}}\leq$ 2.1, 
where $\mathrm{R_{eff}}$ denotes the radius enclosing 50\% of the total galaxy luminosity). 
The thick-orange curve shows a single S\'ersic fit to the central luminosity excess owing to the bulge. 
It can be seen that its isophotal radius \rbulge\ at $\mu=24$ \sbb\ (dashed-dotted vertical line) 
determined from a single-S\'ersic model closely matches the one read off the figure for the case of 
full bulge-bar-disk decomposition.}\label{decomposition} 
\end{figure}
\end{center}

The bulge radius \rbulge\ was determined at an extinction-corrected surface brightness level $\mu_{\rm lim}$ of 24 \sbb\ by fitting a S\'ersic model 
to the interactively selected central luminosity excess of SDSS $r$ SBPs with our surface photometry code iFit (Breda et al., in prep.).
Additionally, a full image decomposition into bulge and disk, and whenever necessary a bar, was carried out with iFit, IMFIT \citep{Erwin2015} and GALFIT \citep{Peng2010} in order to estimate the dependence of \rbulge\ on different codes and profile fitting schemes (see Fig.~\ref{decomposition} for an illustrative example). The latter approach was found to yield a reduction of the radius of the bulge by on average 8\%, thereby leading to only small differences ($\la$10\%) in the quantities inferred for it from spectral modeling (cf. Sect.~\ref{SpectralModeling}), as compared to those within the \rbulge\ from single S\'ersic fits. Given the mean bulge diameter of 12\farcs2$\pm$4\arcsec\ in our sample, and based on simulations, it was found that profile smearing with the typical point spread function (PSF) of SDSS $r$ band data ($\sim$1\farcs3) has a negligible effect on \rbulge\ determinations.
In the following analysis, we refrain from a preliminary photometric subdivision of LTG bulges into CBs and PBs on the basis of the S\'ersic index $\eta$, 
given its unclear physical meaning (cf. Sect.~\ref{intro}). 
This is also because the best-fitting S\'ersic model parameters are sensitive to the details of the SBP decomposition, 
in particular on the modeling and subtraction of the underlying disk \citep[e.g.,][]{P96a,Noeske03}, and eventually the bar \citep[][]{Men08,Breda14}.
This is illustrated on the example of \object{NGC 0776} (Fig.~\ref{decomposition}): it can be seen that automated fitting of a pure exponential to the down-bending (type~ii) profile of the disk would overestimate its central surface brightness and underestimate its scale length $\alpha$. This would then lead to an underestimation of the excess emission from the bulge, and a false determination of its total magnitude, isophotal radius and S\'ersic parameters.
A meaningful approach in this case would be, either to model the disk with a modified exponential distribution 
\citep[for example, the fitting function proposed in][]{P96a}, or to interactively select and fit the inner exponential part of the disk
(1 $\leq$ \rr/\reff $\leq$2.1), as was done here (light-blue line in Fig.~\ref{decomposition}). 
Another salient feature of the SBP of \object{NGC 0776} is a weak bump at 20 $\la$ $\mu$ ($r$ \sbb) $\la$21 that reflects the emission from a bar, being well 
visible on the SDSS true-color image composite (lower-left panel of Fig.~\ref{classes}); its neglect in 1D/2D decomposition could systematically impact 
S\'ersic fits to the excess emission above the disk (open blue crosses), which encompasses the more extended lower-surface brightness (LSB) "wings" 
from the bar. As pointed out in \citet{Breda14}, this could lead to an overestimation of $\eta$ and possibly move a PB into the locus of CBs, eventually also increase the scatter in any relation between $\eta$ and other galaxy parameters (for example, bulge $\sigma_{\star}$, $M_r$, or mean surface brightness).
With these considerations in mind, and to ensure that the evolutionary and spectroscopic properties of bulges (cf. Sect.~\ref{SpectralModeling} and \ref{BPT}, respectively) are obtained within a radius based on a uniform, clear-cut definition and without strong prior assumptions on the photometric structure of LTGs, 
we took the simpler approach of determining \rbulge\ from fitting a single S\'ersic model to the central luminosity peak of the bulge upon visual inspection of the morphology and $g$--$i$ color maps of our LTG sample. 

\subsubsection{Spectral modeling of CALIFA IFS data \label{SpectralModeling}}
Spectral modeling of low resolution ($\lambda/\Delta\lambda\sim 6.5$ at $\sim$5000 \AA) CALIFA IFS data taken with the Potsdam Multi-Aperture Spectrometer \citep[PMAS;][]{Roth05} in its PPaK mode \citep[][]{Verh04,Kelz06} with the V500 grating, and reduced as described in \citet[][and references therein]{GB15CALIFA} was carried out with our pipeline {\sc Porto3D} \citep[][for details]{P13,G16}. 
Spectral fits were computed with the population spectral synthesis (PSS) code \Starlight\ \citep{Cid05} in the spectral range 
between 4000 $\AA$ and 6800 $\AA$ using a library of 152 simple stellar population (SSP) spectra. 
This library (hereafter Z4) comprises SSPs from \citet{BruCha03} for 38 ages between 1 Myr and 13 Gyr for four stellar metallicities 
(0.05, 0.2, 0.4 and 1.0 $Z_{\odot}$), referring to a Salpeter initial mass function and Padova 2000 tracks.
In order to evaluate the robustness of the results, the PSS modeling was repeated for a subset of the data using combined V500 and V1200 (COMB)\footnote{We note that CALIFA DR3 provides COMB data for 97 galaxies from our sample} CALIFA data covering the spectral range between 3700 $\AA$ and 7300 $\AA$. These runs yielded differences of $\la$ 0.2 dex in \mstar, that is within the typical uncertainties expected from PSS modeling with \Starlight\ \citep{Cid14}.

Additionally, the spaxel-by-spaxel modeling 
was repeated in the spectral range between 3900 $\AA$ and 6900 $\AA$ with a SSP library that is identical to Z4 in terms of age coverage except for being supplemented by SSPs with a metallicity of 1.5 \zsun\ (hereafter Z5; 190 elements). A comparison of the results obtained with the Z5 and Z4 SSP base has shown that the global trends between \dmb\ and other quantities considered in this study (stellar mass, surface density, age and metallicity; cf. Fig.~\ref{res}) remain unaltered albeit systematic differences between individual determinations, which presumably reflect the age-metallicity degeneracy \citep[AMD;][]{Worthey94}.
As expected, the Z5-based analysis yields a super-solar metallicity (up to $\sim$1.5 \zsun) for high-\mstar\ bulges at a simultaneous reduction (by up to $\sim$2 Gyr) of \tottmass, whereby differences in \mstar\ and \tsstar\ typically do not exceed 0.2-0.3 dex. As for \dmb\ (Sect.~\ref{RemoveYoung}) the difference between Z5- and Z4-based determinations is rather small ($\pm$0.5 mag) yet systematic, with a weak tendency for a decrease of \dmb\ in high-\mstar\ bulges and vice versa. 
Notwithstanding this fact, the overall robustness of \dmb\ and of its correlation with other properties, despite the notorious AMD, 
is reassuring and underscores its significant potential as a handy proxy of the physical and evolutionary characteristics of LTG bulges.

Whereas the main results from the Z5-based analysis are supplied in the Appendix (Fig.~\ref{Z5-res}) for the disposal of the reader and 
the sake of completeness, we adopt in the following the determinations based on the Z4 SSP library. 
Even though this choice might entail a potential saturation of \totzmass\ at \zsun, it may be expected that, in the presence of the AMD, a narrowing-down of the SSP metallicity space has the advantage of tighter estimates on \tottmass, which is the main focus of this study. Aside from that, it might be conjectured that the occupancy of the full available metallicity space in Z5 fits is partly driven by the mathematical/numerical foundation of state-of-the-art PSS codes, thus in itself no compelling evidence for a significantly over-solar mass-weighted stellar metallicity. 
For instance, whereas the light-weighted \totzlight\ reflects the young and presumably more metal-enriched stellar component (in agreement with super-solar determinations for massive galaxy spheroids from, for example, luminosity-weighted Lick indices), the mass-weighted \totzmass\ primarily reflects the older, high mass-to-light ratio stellar component that likely has had less time for its chemical self-enrichment.
An inequality \totzmass $\leq$ \totzlight\ appears therefore conceivable from the evolutionary point of view.
More generally, it should be kept in mind that the AMD and other potential sources of degeneracy 
\citep[e.g., between metallicity and $\sigma_{\star}$, cf.][]{Koleva08} 
within the complex topology of non-linearly coupled parameters in current PSS models 
have not been fully addressed so far, which makes a conservative limitation of the \zstar\ parameter space an admissible option.
\begin{figure}[htpb] 
\includegraphics[width=1.06\linewidth]{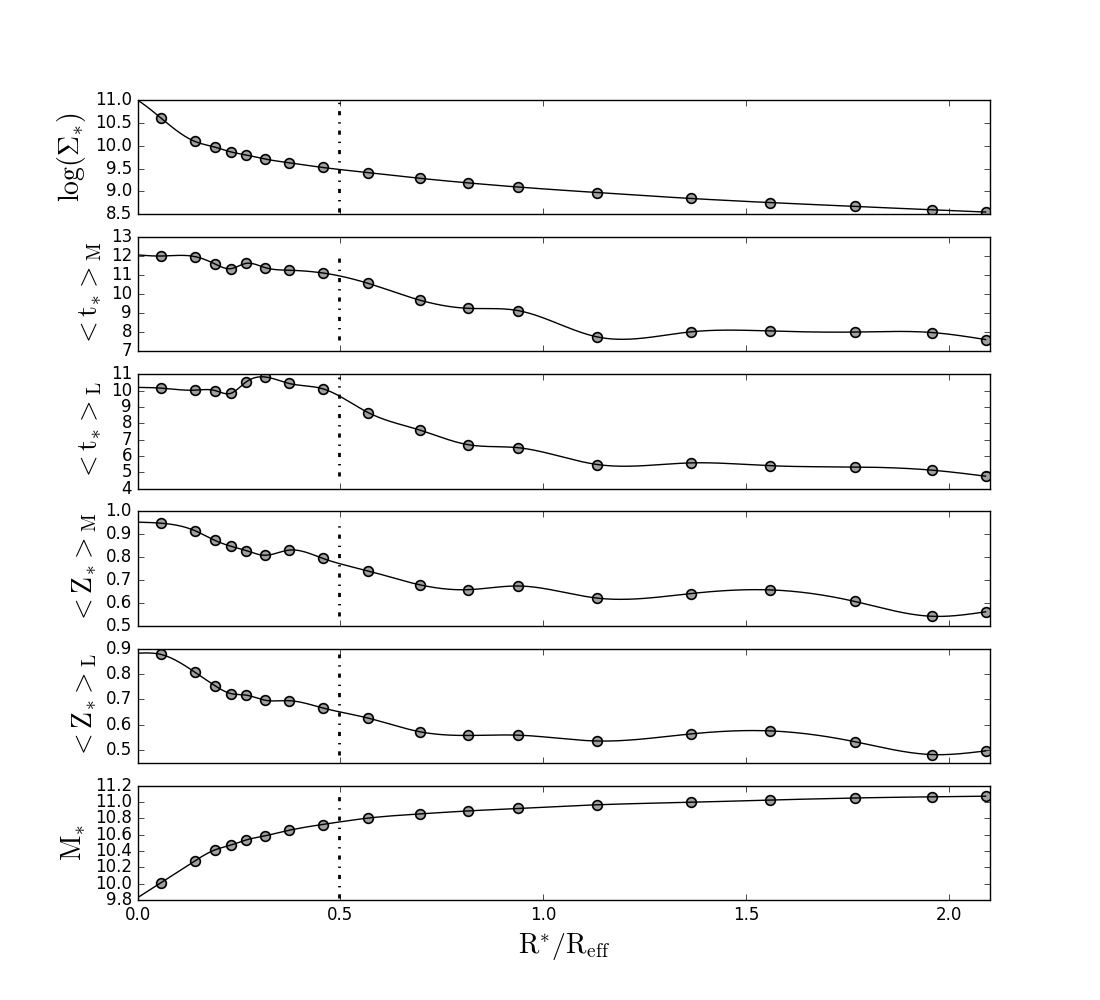}
\caption[Radial profiles of various physical quantities for NGC 0760]{Radial profiles for various quantities obtained for \object{NGC 0776} 
with the \brem{isan} technique and auxiliary codes for spline interpolation and statistical analysis. From top to bottom the panels show: 
i) a logarithmic representation of the stellar surface density \tsstar\ in \msun/kpc$^{2}$, ii\&iii) the mass- and light-weighted stellar age (\utmass\ and \utlight, respectively) in Gyr, iv\&v) the mass- and light-weighted stellar metallicity (\totzmass\ and \totzlight, respectively) in \zsun, 
and vi) the enclosed present-day stellar mass \mstar\ as a function of the photometric radius \rr\ normalized to the effective radius \reff.
The bulge radius \rbulge\ is depicted by the vertical dashed line.}\label{prof} 
\end{figure}

{\sc Porto3D} computes several quantities of interest, including the present-day and ever formed \mstar\ (\msun), \tottmass\ (Gyr) and metallicity \totzmass\ ($Z_{\odot}$), and their luminosity-weighted values (\tottlight\ and \totzlight), the stellar surface density \tsstar\ (\msun/kpc$^{2}$),
the time \thalf\ when 50\% of the present-day \mstar\ was in place, the light and mass-fraction of stellar populations younger than 0.1, 1 and 5 Gyr, as well as emission-line fluxes and equivalent widths (EWs), and stellar and ionized-gas velocity maps \citep[see][for details]{G16}. 
Here we limit the discussion to mass-weighted values, since they are robust against young ($\la$0.1 Gyr) stellar populations that typically dominate the light despite their very low \mstar\ fraction.

Spatially resolved maps of the aforementioned quantities were in turn converted into radial profiles using an adaptation of the 
isophotal annuli (\brem{isan}) surface photometry technique by \citet{P02}. The key feature of this method \citep[e.g.,][]{Kehrig12,P13,G16} 
consists in the computation of statistics within logarithmically equidistant isophotal zones defined from a reference image -- in this case, the emission-line-free pseudo-continuum between 6390 $\AA$ and 6490 $\AA$, segmented into 18 isophotal zones. The latter closely trace the galaxy morphology at all surface brightness levels, without the prior assumptions on galaxy structure commonly made in 1D/2D surface photometry techniques (approximation of a galaxy as due to superposition of axis-symmetric components, or profile derivation within elliptical annuli with constant ellipticity and position angle), this way permitting accurate determination of SBPs and color profiles for irregular galaxies \citep[see also, e.g.,][]{Noeske03}. 
Figure~\ref{prof} shows an example of radial profiles, after normalization\footnote{We note that the normalization of radial profiles in Fig.~\ref{prof} to $r_{\rm n}=$\rr/\reff\ is for the sake of illustration only. Such a normalization is omitted in the forthcoming discussion because it could bias studies of the metallicity and star formation history (SFH) of the bulge and disk component of LTGs. This is because $r_{\rm n}$ is coupled with the B/T ratio: for instance, from Fig.~8 of \citet{P06} it follows that the \reff\ of a bulgeless LTG corresponds to 1.7 exponential scale lengths $\alpha$ of the disk, whereas in a bulge dominated LTG with a B/T=3/4 the \reff\ shrinks to $\sim$0.5\,$\alpha$.     
Therefore, stacking or inter-comparison of $r_{\rm n}$-normalized profiles in an attempt to systematize radial trends in, for example, age, SFR and specific SFR (sSFR) 
in the bulge and disk of LTGs can lead to averaging of determinations within the bulge component of higher-B/T LTGs with those in the disk component of lower-B/T LTGs.} to \reff.

In turn, \brem{isan} determinations were spline-interpolated to a finer radius step to ensure that average values within \rbulge\ are not biased toward a group
of data points that are eventually densely spaced in radius. The bulge mean stellar age \mtmass\ and metallicity \mzmass\ were obtained as the arithmetic mean of these values, 
and the present-day stellar mass M$_{\star,\mathrm{B}}$ of the bulge was computed by integrating \tsstar\ 
profiles out to \rbulge. 
Additionally, mean values for the bulge were computed by performing statistics directly on 2D maps within the isophote corresponding  
to \rbulge, finding overall a satisfactory agreement with the previous determinations.

Even though all quantities above are weighted by \mstar, being therefore relatively insensitive to the luminosity contribution by the star-forming disk, the fact that \rbulge\ extends in some cases (e.g., \object{NGC 0776} in Fig.~\ref{decomposition}) far into the latter, and the  
bulge line-of-sight contribution sharply decreases beyond the bulge \reff, calls for an evaluation of a possible contamination by the disk.
A correction for the latter would in principle be possible, if its properties (\tsstar, \utmass\ and \totzmass) 
beneath the bulge could be constrained with sufficient accuracy, or at least coarsely from their estimated luminosity fraction within \rbulge, or using hybrid spectro-photometric decomposition techniques \citep[e.g.,][]{Johnston17}. These approaches, however, rely by necessity on simplifying assumptions on the photometric structure of LTGs, the most important of which being that the disk preserves its exponential slope at $R^{\star}$ $>$ \rbulge\ all the way to its center, and has throughout a zero radial gradient in age, mass-to-light ratio and metallicity. 
As these assumptions are controversial or incompatible to observations \citep[e.g.,][]{P96a,Noeske03,SB14,Tis16}, no attempt was made for a spectrophotometric disk subtraction, consistently with the approach taken in the photometric analysis (Sect.~\ref{SurfPhot}).
However, a series of tests made within the central portion (\rr$\leq$3\farcs3) of bulges, where contamination by the disk 
is minimal, has shown that mass-weighted quantities inferred therein are in good agreement with those within \rbulge\ 
(cf. Fig.~A.1 in Sect.~\ref{A-check1}). This indicates that the luminosity contribution by the star-forming disk does not appreciably impact determinations 
of \mtmass\ and \mzmass\ in Fig.~\ref{res}.
\subsubsection{Post-processing of the spectral synthesis output with {\sc RemoveYoung} \label{RemoveYoung}} 
{\sc RemoveYoung} \citep[\RY;][]{GP16-RY} is a tool intended to the post-processing of the population vector (PV; i.e., the best-fitting combination of fractional contributions of individual SSPs to the galaxy mass) obtained by modeling a spectrum with a PSS code (e.g., {\sc Starlight} in this case). \RY\ permits removal from a PV of the contribution from SSPs younger than an adjustable age cutoff $t_{\rm cut}$ and computation of the spectrum, magnitudes in different filters (e.g., SDSS $u$, $g$, $r$, $i$, $z$) and stellar mass of the residual older stellar component. 
In particular, spaxel-by-spaxel application of \RY\ to the spectral synthesis output from {\sc Porto3D} allows to strip off IFS data cubes from, 
for example, the young ionizing stellar component ($t_{\rm cut}\sim 30$ Myr) and produce synthetic images of the underlying older stellar background in various photometric bands 
(\RY\ convolves the residual spectral energy distribution (SED) with the filter transmission functions).
In the framework of this study, \RY\ was applied spaxel-by-spaxel to the LTG sample for eight $t_{\rm cut}$ values (0.03, 0.1, 0.3, 1, 3, 5, 7 and 9 Gyr). 
The $\mu$ and \tsstar\ maps from \RY\ were in turn converted into 1D radial profiles as described in Sect.~\ref{SpectralModeling}.

Figure~\ref{classes} illustrates the application of \RY\ on three LTGs from our sample. The synthetic $r$ band SBPs 
for the eight adopted age cutoffs are shown color coded. SBPs labeled "$r$ SDSS" were obtained for a $t_{\rm cut}=0$ Gyr, that is, through 
convolution of the {\sl observed} IFS data cubes with the SDSS $r$ band transmission curve.  
For comparison, we overlay the $r$ band SBPs computed directly from SDSS images (light-gray curves labeled "OBS $r$ SDSS"), which due to their better
resolution (FWHM$\sim$1\farcs3 as compared to $\sim$2\farcs6 for CALIFA IFS data) better trace the central luminosity peak of the bulge. 
\linebreak
It can be seen that whereas all three LTGs show a strong surface brightness dimming with increasing $t_{\rm cut}$ in their disks, this is not necessarily 
the case for their bulge component (vertical gray line). For example, suppression of stellar population of a successively higher age results in a 
roughly uniform dimming both in the disk and bulge of \object{IC 0776}, whereas it has practically no effect on the bulge surface brightness of 
\object{NGC 0776}.
\begin{center}
\begin{figure} 
\includegraphics[width=1\linewidth]{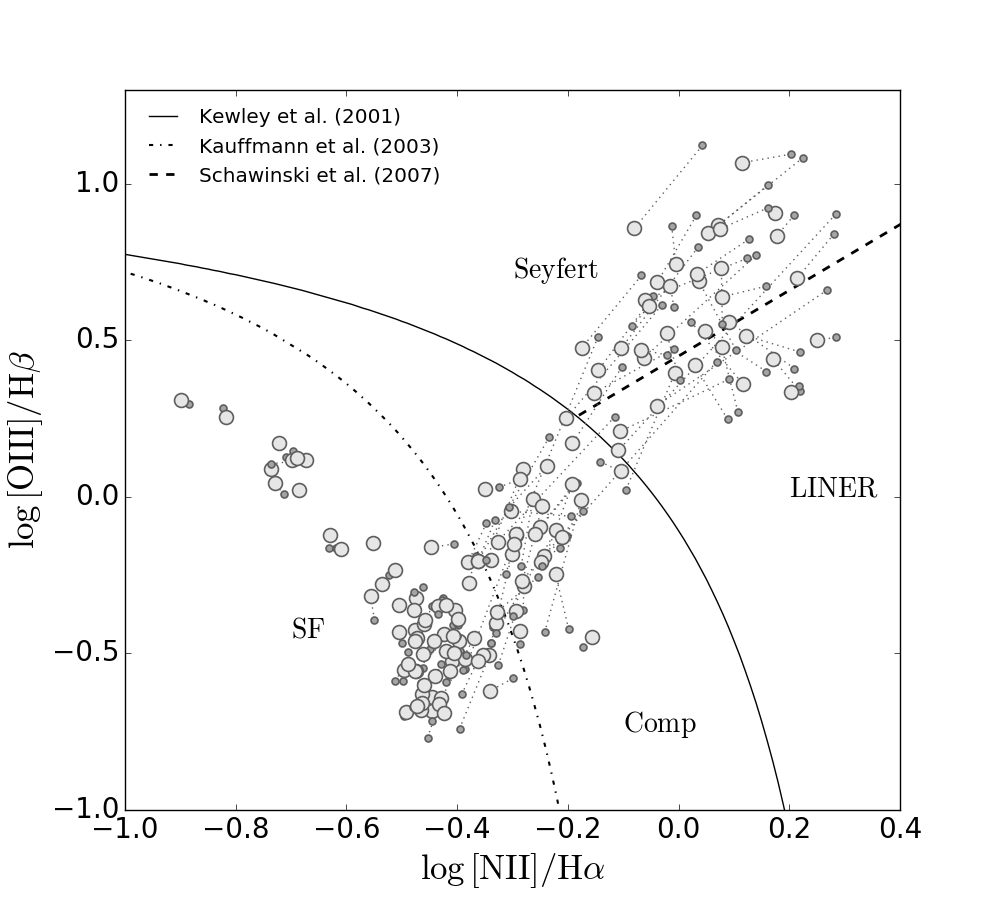}
\caption[Comparison of BPT determinations with different methods]{
Comparison of diagnostic emission-line ratios after \citet{BalPhiTer81}, as obtained within a 3\arcsec\ aperture (small circles; method \brem{a}), 
connected by dotted lines with determinations from method \brem{b} (mean ratios within \rbulge\ from radial [O{\sc iii}]/H$\beta$ and [N{\sc ii}]/H$\alpha$ 
profiles; big circles). 
Data based on method \brem{c} (luminosity-weighted determinations using integral [O{\sc iii}]${5007}$, H$\beta$, [N{\sc ii}]${6584}$ and H$\alpha$ line fluxes within \rbulge) and error bars are omitted for the sake of better visibility.
The loci of Seyfert, LINERs and Composites, and that corresponding to photoionization by SF are demarcated following \citet[][dashed-dotted curve]{Kauffmann03b}, \citet[][solid curve]{Kewley01} and \citet[][dashed line]{Schawinski2007}.}
\label{BPT-comparison} 
\end{figure}
\end{center}

Based on the output from \RY, we computed the difference $\mu_{\rm 0\,Gyr}-\mu_{\rm 9\,Gyr}$ where $\mu_{\rm 0\,Gyr}$ and $\mu_{\rm 9\,Gyr}$ denote, respectively, the synthetic $r$-band SBP of each LTG for a $t_{\rm cut}$ of 0 Gyr and 9 Gyr. The arithmetic average of this profile within \rbulge\ is referred to in the following as \dmb\ (mag).
For instance, a \dmb=0 mag corresponds to the case where stellar populations with age $\geq$9~Gyr entirely dominate the  
$r$-band surface brightness within \rbulge, whereas a \dmb\ of --2.5 mag translates into a contribution of 10\% by this old stellar component.
Even though a correlation between \dmb\ and broadband colors appears plausible, it should be born in mind 
that these two quantities have a different definition, and the former (as well as any other quantity comparing $\mu$ for two different $t_{\rm cut}$'s, for example, 0.3 and 1 Gyr) yields a stronger age diagnostic, since, at variance to colors, in principle permits complete suppression of stellar populations younger than a given age.
As we shall argue next, \dmb, introduced and analyzed for the first time here, offers a simple, distance- and formally
extinction-independent proxy to the evolutionary and physical properties of LTG bulges
\footnote{We note that the stellar extinction in the bulge component of our LTG sample is relatively low, with a mean value of $A_V=0.3\pm0.18$ mag.
\dmb\ can be readily obtained by post-processing a PSS fit to any spectroscopic data set
(e.g., single-fiber spectroscopy from the SDSS and GAMA surveys) with the publicly available version of \RY\ 
(cf. www.iastro.pt/research/tools/RemoveYoung.html).}.
\subsection{Spectroscopic classification of LTG bulges \label{BPT}}
Another aim of this study is the exploration of gas excitation mechanisms in LTG bulges across their relevant range in \mstar\ and \tsstar\  using a combination of methods that allow to quantify aperture biases in luminosity-weighted emission-line ratios.
To this end, emission-line maps were determined by {\sc Porto3D} through spaxel-by-spaxel subtraction of the best-fitting stellar model from the input spectrum. The flux of the four emission lines ([O{\sc iii}]${5007}$, H$\beta$, [N{\sc ii}]${6584}$ and H$\alpha$) used for spectroscopic classification after \citet[][hereafter BPT]{BalPhiTer81} were then determined from stellar continuum-subtracted maps. An accurate starlight subtraction is crucial to the spectroscopic classification, inter alia because in areas with weak nebular emission, hydrogen Balmer lines embedded within broader stellar absorption profiles are eventually strongly underestimated when measured with standard line fitting techniques. As pointed out in \citet{Pet11}, this could in turn artificially increase the [O{\sc iii}]/\hb\ ratio, moving a source upward on BPT diagrams and eventually prompting its erroneous classification as a Seyfert. 

BPT ratios for our LTG bulges were determined with three different methods.  The first one (method \brem{a}) 
simulates SDSS measurements within a 3\arcsec\ aperture centered on the maximum of the emission-line-free pseudo-continuum maps. This aperture diameter yields a good match to the angular resolution of CALIFA data and is generally much smaller than \rbulge\ 
(cf. Sect. \ref{SpectralModeling} \& Sect.~\ref{A-check1}), thereby minimizing dilution of possible spectroscopic signatures from an Active Galactic Nucleus (AGN) by circumnuclear SF.
Additionally, BPT ratios were determined by averaging out to \rbulge\ spline-interpolated [O{\sc iii}]/H$\beta$ and [N{\sc ii}]/H$\alpha$ ratios 
within \brem{isan} (method \brem{b}), as well as from the integrated [O{\sc iii}]${5007}$, H$\beta$, [N{\sc ii}]${6584}$ and H$\alpha$ line fluxes 
within \rbulge\ (method \brem{c}).
Whereas determinations with method \brem{a} are most sensitive to a central AGN, the contrary is the case for the area-weighted BPT ratios from method \brem{b}. This is because the latter are only weakly dependent on the luminosity contribution from a bright central point source, which, even if dominating the total nebular luminosity within the bulge, would affect radial 
log\,[O{\sc iii}]/\hb\ and log\,[N{\sc ii}]/\ha\ profiles only locally (out to \rr$\sim$FWHM), thereby having little influence on their
mean value within \rbulge. Method \brem{c}, on the other hand, yields purely luminosity-weighted BPT ratios within \rbulge\ and simulates the (idealized) situation where 
the SDSS fiber precisely matches the isophotal bulge diameter of a LTG.

Differences between method \brem{b} and \brem{c} were found to be generally small ($\la$0.3 dex) and in most cases compatible to those obtained 
with method \brem{a}, therefore not globally altering the spectroscopic classification of the analyzed bulges, 
in particular for sources classified as SF. 
However, as apparent from the upper-right part of Fig.~\ref{BPT-comparison}, there is a tendency for method \brem{b} (large circles) 
to move determinations within a 3\arcsec\ aperture (method \brem{a}, small circles) from the locus of LINERs \citep{Heckman1980} and Seyferts 
into the locus of Composites.

This trend, which can be attributed to dilution of central AGN/LINER emission by circumnuclear SF, is in accord with the conjecture by \citet{Gomes16b} that the upper-right "wing" delineated by SDSS galaxies on the BPT diagram is partly due to aperture effects and consistent with an inside-out galaxy formation (or, SF quenching) scenario.
It is also interesting in this context that \citet{Igl16} document from analysis of CALIFA IFS data that integrated [O{\sc iii}]/H$\beta$ and [N{\sc ii}]/H$\alpha$
ratios for local LTGs can differ from single-fiber SDSS determinations by up to $\sim$0.3 dex. 

It is worth noting that, whereas Seyfert or Composite BPT ratios imply that an AGN dominates or substantially contributes to the gas excitation, 
the role of accretion-powered nuclear activity in bulges classified as LINERs is less clear. Traditionally, LINER emission-line ratios in massive,  
high-\tsstar\ spheroids (early-type galaxies and bulges) were ascribed to a diffuse floor of photoionization powered 
by the hard radiation field from hot evolved ($\geq 10^8$ yr) post-asymptotic giant branch (pAGB) stars  
\citep[e.g.,][]{tri91,bin94,macc96,sta08,cid10,cid11,Sarzi10,YanBlanton2012}. However, the pAGB photoionization hypothesis is valid only as long as the observed EW(\ha)$_{\rm obs}$ does not exceed $\sim$3 $\AA$, since this is the maximal, nearly metallicity-independent value predicted by zero-dimensional (0D) ESS models for an old, instantaneously formed stellar population, provided that case~B recombination applies \citep[e.g.,][]{cid11,G16}. 
More specifically, as pointed out in \citet[][hereafter P13]{P13}, whereas an EW(\ha)$_{\rm obs}\leq 3$ $\AA$ is 
a necessary condition for the pAGB photoionization hypothesis to be tenable, it is in itself no compelling evidence against an AGN\footnote{ 
This is because in a triaxial geometry, dilution of the intrinsic (nuclear) EW(\ha)$_{\rm nuc}$ by the stellar background along the line of sight implies an EW(\ha)$_{\rm nuc}$ $\geq$ EW(\ha)$_{\rm obs}$, with equality between these quantities representing a special case.
Actually, the spatial anti-correlation between emission-line EWs and \tsstar\ observed in many star-forming galaxies nicely illustrates this effect \citep{P02}, and together with the considerations above calls attention to the fact that interpreting projected observables 
and byproducts thereof (e.g., colors, EWs, Lick indices and ages, SFHs, metallicity enrichment histories, respectively) using 
predictions from standard 0D evolutionary synthesis models is not a straight forward task.\label{footnote4}}.

Other interpretations for the origin of LINER emission involve fast shocks \citep[e.g.,][]{DopitaSutherland1995,Allen2008} and gas excitation  
by a radiatively inefficient low-luminosity AGN \citep[e.g.,][]{Ho1999}. P13 argue that even a strong AGN can not be ruled out in the LINER nuclei of many early-type galaxies (ETGs). This is because the Lyman continuum (Ly$_{\rm c}$) photon escape fraction in the centers of these systems can reach values $\ga$0.9, which implies that the bulk of ionizing radiation from a putative AGN escapes without being locally reprocessed into nebular emission. Ly$_{\rm c}$ photon escape, in conjunction with the EW dilution effect provides therefore an ansatz for understanding why many ETGs with clear evidence for a prodigious energetic output from an AGN (e.g., radio- and even X-ray jets, as in the case of the LINER ETG M87 in the Virgo galaxy cluster) show weak, if any at all, nebular line emission. As conjectured in P13, the typically LINER BPT ratios in these massive galaxy spheroids may witness a situation where tenuous gas -- permitting due to its very low density a high Ly$_{\rm c}$ escape fraction -- is exposed to the hard radiation field from an AGN and the diffuse post-AGB component. Indeed, the Ly$_{\rm c}$ escape fraction is anti-correlated with EW(\ha) (cf. their Fig.~2) and invariably exceeds 0.5 in LINER ETG nuclei.
Evidently, the same association between LINER emission and extensive Ly$_{\rm c}$ photon escape from a virtually gas-evacuated 
high-\tsstar\ stellar spheroid is also conceivable for massive LTG bulges hosting an AGN.
On the basis of such considerations we do not exclude in the discussion in Sect.~\ref{dis} that LTG bulges falling in the LINER locus of BPT diagrams could host significant accretion-powered nuclear activity.

\section{Results \label{res1}}
In this section we provide an overview of the main results obtained,
laying emphasis on the relation between \dmb\ and the evolutionary and physical characteristics of LTG bulges in our sample. 
Additionally, we examine the variation of the bulge-to-disk age and metallicity contrast as a function of LTG mass, as well as 
the variation of the dominant gas excitation mechanisms along the bulge mass and age sequence.

\subsection{Physical and evolutionary properties of LTG bulges vs. \dmb\ \label{sect3-a}}
\begin{center}
\begin{figure}[htpb] 
\includegraphics[width=1\linewidth]{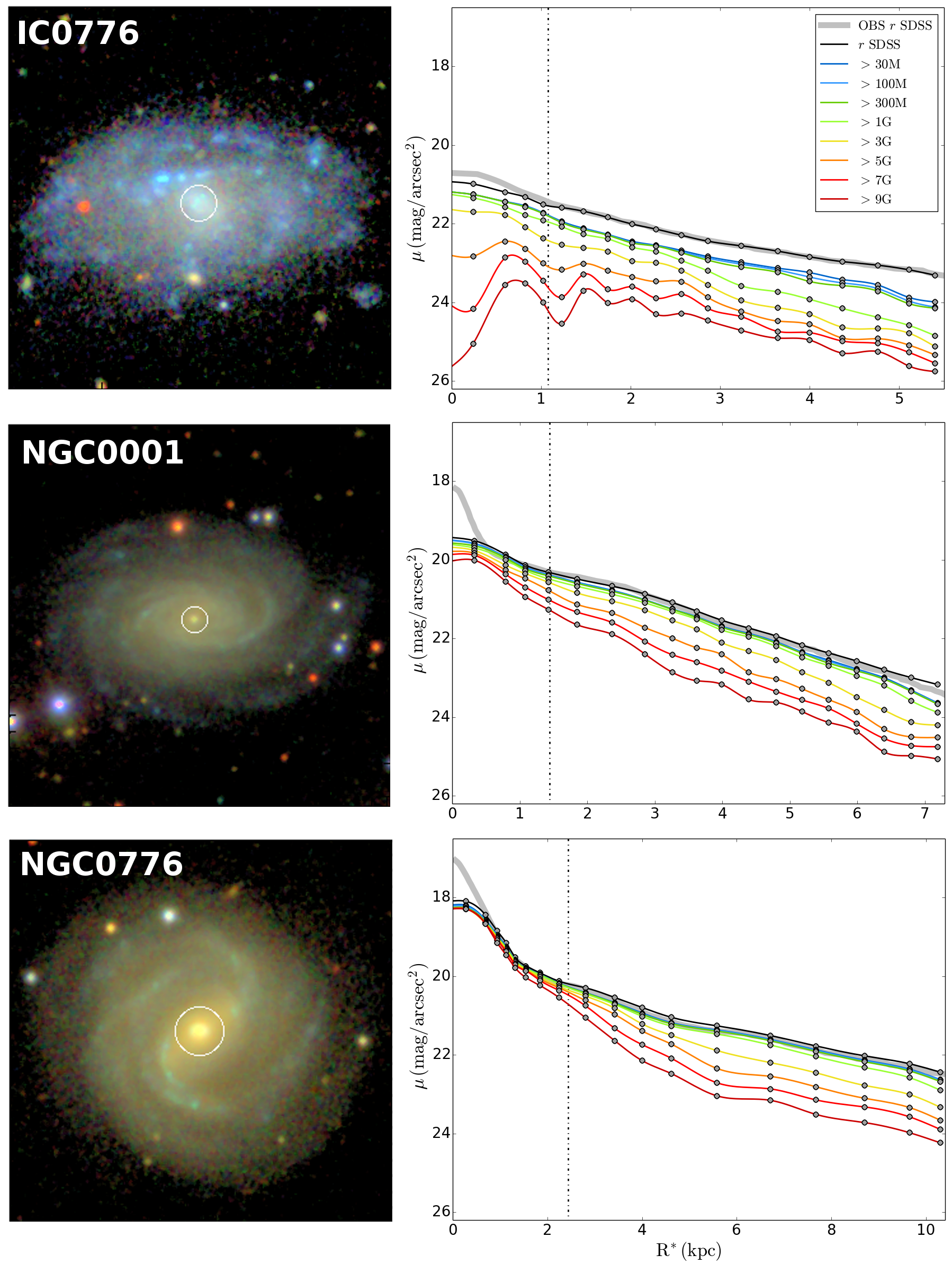}
\caption{SDSS true-color images and SBPs (left- and right-side panel, respectively) for three LTGs 
illustrating the prominence of the bulge relatively to the disk for the three \dmb\ intervals tentatively defined in Sect.~\ref{res1}:   
IC 0776 (log\,\mstotal = 9.58; interval~\brem{iA}), NGC\ 0001 (log\,\mstotal = 10.99; interval~\brem{iB}) and NGC 0776 (log\,\mstotal = 11.09; interval~\brem{iC}).
Synthetic SBPs computed through convolution of the observed IFS data with the SDSS $r$-band filter transmission curve are shown in black, and those 
after removal with \RY\ of stellar populations younger than 0.03, 0.1, 0.3, 1, 3, 5, 7 and 9 Gyr with the color coding in the upper-right panel. Thick-gray curves depict $r$-band SBPs computed from SDSS images and used for fitting a S\'ersic model to infer the bulge radius \rbulge\ (dashed vertical lines). The circles overlaid with SDSS images depict the bulge diameter.}\label{classes} 
\end{figure}
\end{center}

Figure~\ref{classes} illustrates three characteristic snapshots along a sequence of increasing galaxy \mstar\ and bulge \dmb: whereas 
all three LTGs shown display a significant ($>$1.5 mag) surface brightness dimming with increasing $t_{\rm cut}$ in the disk (\rr $\geq$ \rbulge), implying its continued growth over the past 9 Gyr, their bulge \dmb\ spans a broad range between $\sim$0 mag and $\sim$--4 mag, which translates, respectively, into a $r$-band luminosity fraction between $\sim$0\% and $\sim$97\% by stars younger than 9 Gyr.

A salient feature of almost bulgeless LTGs (e.g., \object{IC 0776}) with a\linebreak \dmb\ $\leq$ --1.5 mag 
(hereafter, \dmb\ interval \brem{iA}; 34 galaxies) is a roughly uniform dimming in $\mu$ with increasing $t_{\rm cut}$ both in the bulge and the disk, which is consistent with a nearly homologous growth of \tsstar\ throughout the galaxy's extent. 
To the contrary, systems like \object{NGC 0776} (\dmb\ $\geq$ --0.5 mag; interval~\brem{iC}: 43 galaxies) show strong recent evolution only in their disks, whereas their \dmb\ documents a dominant old stellar population in the bulge with no appreciable SF occurring therein over the past 9 Gyr. As for LTGs in the intermediate range of \dmb\ (--1.5 mag to --0.5 mag; subset~\brem{iB}: 58 galaxies; e.g., \object{NGC 0001}), spectral modeling points to a significant contribution from stars younger than 9 Gyr to the bulge luminosity.

A question next is, how the physical and evolutionary properties of LTG bulges may vary across these three tentatively defined intervals 
in \dmb.

A synopsis of the main results obtained for our sample is given in Fig.~\ref{res} (see also Table~1).
Panel~\brem{a} reveals a nearly linear relation between \dmb\ and mass-weighted stellar age \mtmass\ for the bulge, with the transition from  interval~\brem{iA} (blue dots) to interval~\brem{iB} (green dots) occurring at \mtmass $\sim$ 9 Gyr, and bulges in the interval ~\brem{iC} (red dots) populating the upper-right part of the diagram (\mtmass $\ga$ 11 Gyr). These comparisons show that bulges have a large range in SSP-equivalent ages from $\sim$2 to 13.5 Gyr \citep[cf. e.g.,][]{Peletier07,MoHo06} and metallicities. 

The tight trend between age and \dmb\ in this panel can be approximated \footnote{Linear fits in this section and in Sect.~\ref{ap:z} were computed through $\chi^2$ minimization for both quantities considered, given that none of them is strictly independent. Table~\ref{linfit} in the Appendix additionally lists the coefficients from regression analysis minimizing $\chi^2$ along the abscissa only.} by the relation
\mtmass\ (Gyr) = (12.14$\pm$0.06) + (2.05$\pm$0.04) $\cdot$ \dmb\ (solid line).

Panel~\brem{b} shows that the age of LTG bulges (Gyr) roughly scales with the logarithm of their stellar mass (\msun) as 
\mtmass\ = (2.30$\pm$0.11)$\cdot \log$\,\mbstar\ - (13.44$\pm$1.13).
This implies that old ($\geq$11 Gyr) bulges in the \dmb\ interval \brem{iC} are by $\sim$2 orders of magnitude more massive than 
bulges falling in the \dmb\ interval \brem{iA}. The mean age of the latter was determined to be 6.82$\pm$1.06 Gyr, nearly 5 Gyr lower than that of bulges in the interval \brem{iC} (10.7$\pm$0.4 Gyr), with bulges in the interval \brem{iB} having an average age of 9.80$\pm$0.7 Gyr. 
As apparent from panels~\brem{d}\&\brem{e}, massive old bulges (\brem{iC}) are the most metal enriched and show the highest \sstar, which in all cases exceeds $10^9$ \msun\, kpc$^{-2}$ and reaches up to $\sim 6 \times 10^9$ \msun\, kpc$^{-2}$, whereby the bulge stellar surface density scales as 
$\log$\,\sstar = (4.96$\pm$0.23) + (0.42$\pm$0.02)$\cdot \log$\,\mbstar\ and
$\log$\,\sstar = (7.63$\pm$0.09) + (0.16$\pm$0.01)$\cdot$\mtmass.

From the combined evidence of panels \brem{a}, \brem{b} and \brem{e}, a trend between \dmb\ with bulge mass and surface density is to be expected. Indeed, linear fits to the data yield tight relations of the form 
$\log$\,\mbstar = (10.87$\pm$0.06) + (0.67$\pm$0.04)$\cdot$\dmb\ 
and 
$\log$\,\sstar = (9.56$\pm$0.03) + (0.32$\pm$0.02)$\cdot$\dmb,
suggesting that \dmb\ offers a useful semi-empirical proxy to the evolutionary and physical properties of LTG bulges.

\setlength{\unitlength}{1pt}
\begin{center}
\begin{figure*} 
\begin{picture}(180.4,310.0)
\put(0,0){\includegraphics[width=1\linewidth]{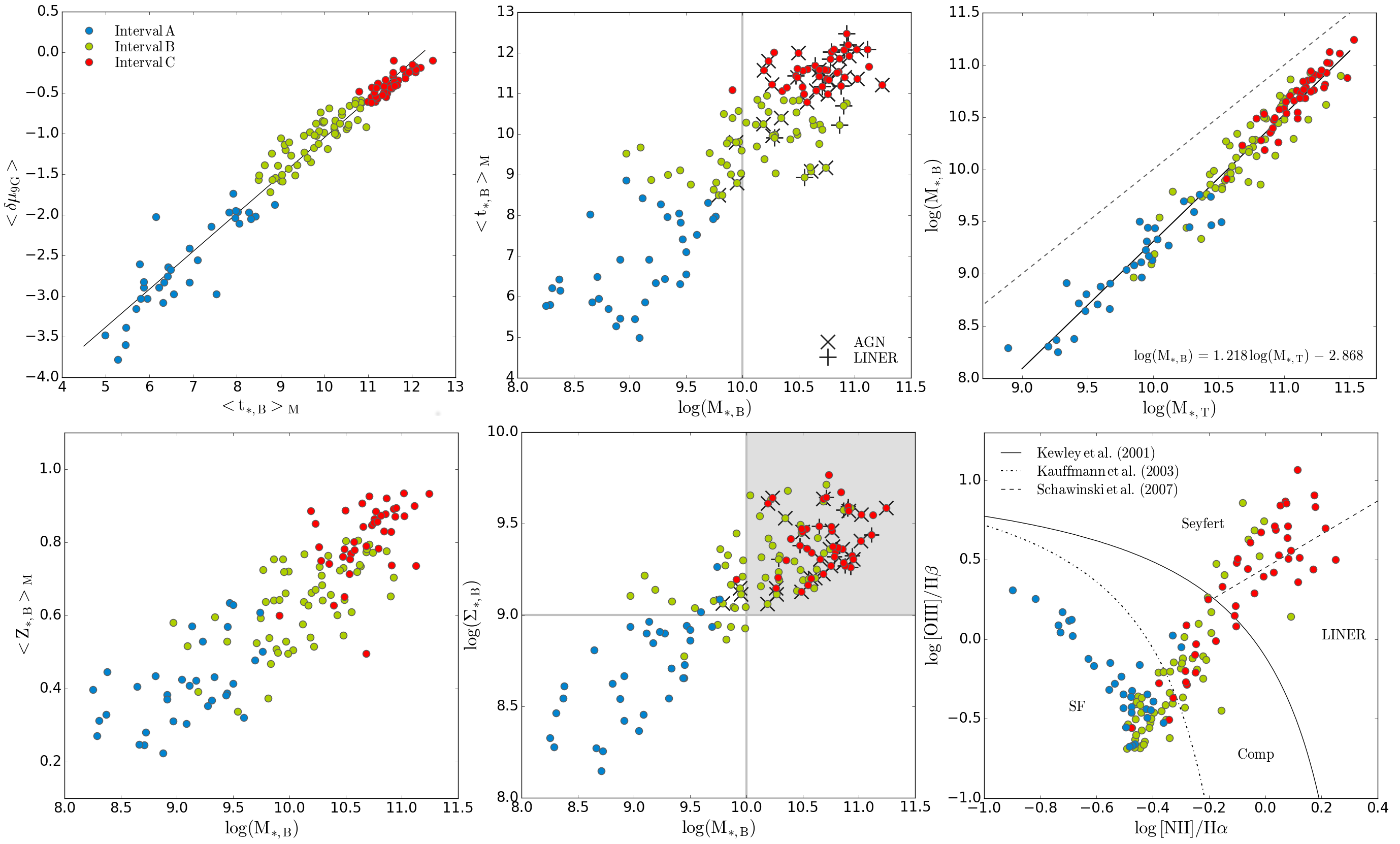}}
\PutLabel{157}{180}{\hvss a}
\PutLabel{200}{296}{\hvss b}
\PutLabel{376}{296}{\hvss c}
\PutLabel{157}{22}{\hvss d}
\PutLabel{330}{22}{\hvss e}
\PutLabel{376}{22}{\hvss f}
\end{picture}
\caption{
\brem{a)} \dmb\ ($r$ mag) vs. mass-weighted stellar age \mtmass\ (Gyr) for the bulge component of our sample LTGs. The solid line shows a linear fit to the data.
\brem{b)} Logarithm of the stellar mass \mbstar\ (\msun) in the bulge vs. mass-weighted stellar age \mtmass\ (Gyr). Galaxies spectroscopically classified as Seyfert and LINER are marked. 
\brem{c)} Total stellar mass \mstotal\ vs. \mbstar. The solid and dashed lines show, respectively, a linear fit to the data and equality between both quantities. 
\brem{d)} Logarithm of \mbstar\ vs. mass-weighted stellar metallicity \mzmass\ (\zsun).
\brem{e)} Logarithm of \mbstar\ vs. logarithm of the mean stellar surface density \sstar\ (\msun\,kpc$^{-2}$) in the bulge. 
The gray-shaded quadrant at log\,\mstar$\geq$10 and log\,\sstar$\geq$9 delineates the parameter space containing 93\% of all Seyfert and LINER galaxies in our sample.
\brem{f)} Spectroscopic classification after BPT within simulated 3\arcsec\ SDSS fibers (method \brem{a} in Sect. \ref{BPT}; cf. Fig.~\ref{BPT-comparison}).
}\label{res}
\end{figure*}
\end{center}      
Finally, within the considered range of stellar metallicities (\zsun/20 -- \zsun), linear fits to \mzmass\ (panel \brem{d}) yield the relations
\mzmass\ (\zsun) = (0.23$\pm$0.01)$\cdot \log$\,\mbstar\ -- (1.60$\pm$0.11) and
\mzmass\ (\zsun) = (0.28$\pm$0.02)$\cdot \log$\,\mstotal\ -- (2.37$\pm$0.15).

It is interesting to note that the age and metallicity of the disk (\rr$\geq$\rbulge) follow similar relations with total stellar mass \mstotal\ (\msun),
as \mdtmass\ = (1.88$\pm$0.12)$\cdot \log$\,\mstotal\ -- (11.85$\pm$1.32)
and 
\mdzmass\ (\zsun) = (0.20$\pm$0.02)$\cdot \log$\,\mstotal\ -- (1.55$\pm$0.16). 

Another issue of interest concerns the contribution of the bulge to the total LTG mass \mstotal.
Panel \brem{c} shows that more massive bulges are hosted by more massive (and luminous) LTGs, following a relation 
$\log$\,\mbstar = (1.22$\pm$0.02) log\,\mstotal\ -- (2.87$\pm$0.25), or, equivalently,
\mbstar/\mstotal\ = (0.13$\pm$0.02)$\cdot$log\,\mstotal\ -- (1.12$\pm$0.17).
The obtained relation (solid line) implies that bulges in the \dmb\ interval \brem{iC} contain on average $\sim$40\% of the total stellar 
mass of a LTG, twice the mass fraction determined for bulges falling in the interval \brem{iA}.
\setlength{\unitlength}{1pt}
\begin{center}
\begin{figure*}[htpb] 
\begin{picture}(180,310)
\put(0,0){\includegraphics[height=11cm]{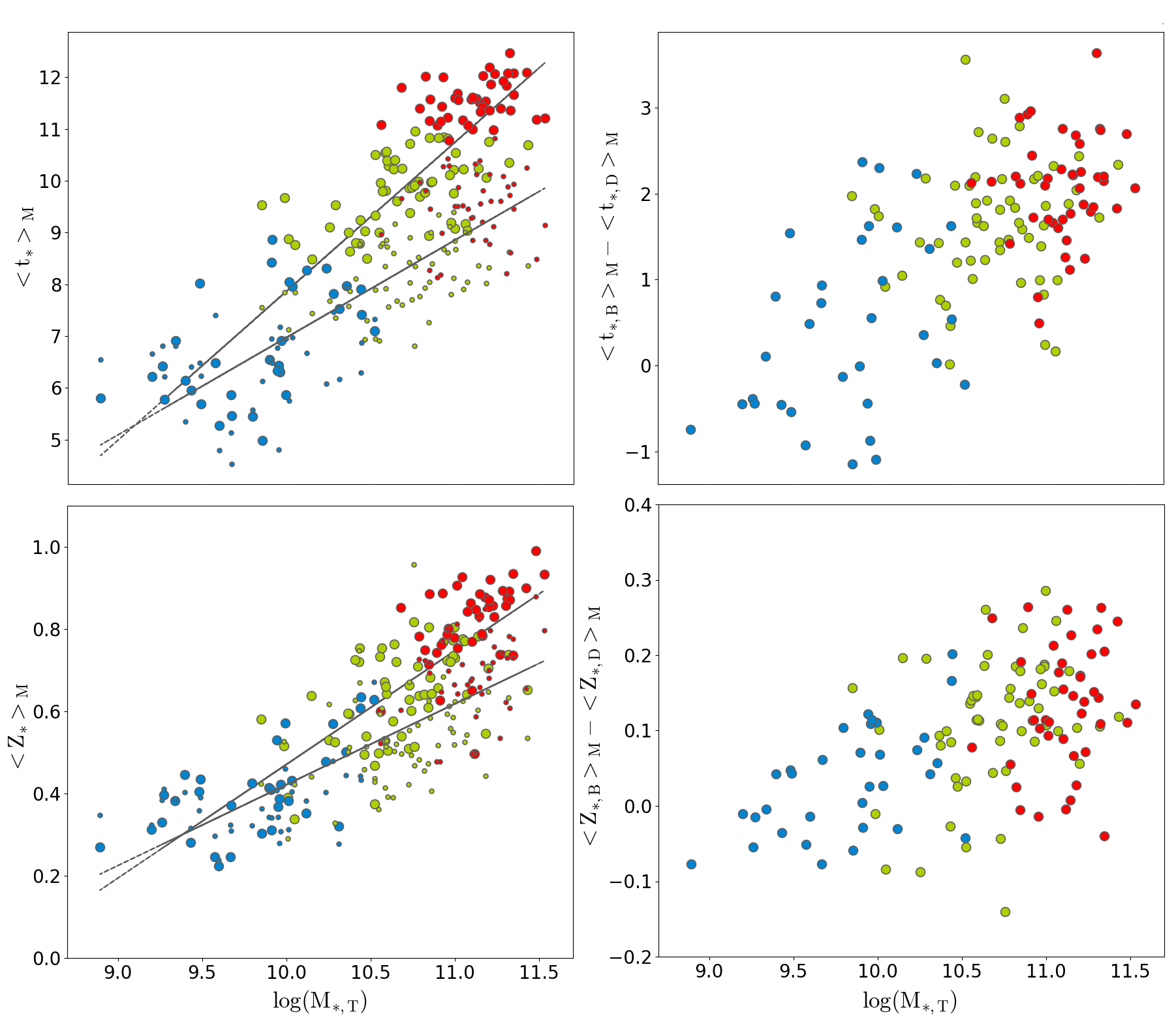}}
\PutLabel{30}{290}{\hvss a}
\PutLabel{30}{146}{\hvss b}
\PutLabel{212}{290}{\hvss c}
\PutLabel{212}{144}{\hvss d}
\PutWin{380}{234}{5cm}{\caption{
\brem{a)} Logarithm of the total stellar mass \mstotal\ (\msun) vs. mass-weighted stellar age \utmass\ (Gyr) of the bulge and disk component (large and small dots, respectively). Linear fits to the data for the bulge and the disk are shown with solid lines. The color coding is the same as in Fig.~\ref{res}. 
\brem{b)} log(\mstotal) vs. mass-weighted stellar metallicity (\zsun) for the bulge and the disk. The layout is identical to that in panel~a.
\brem{c)} log(\mstotal) vs. age contrast \dbdt\ (\mtmass\ -- \mdtmass) (Gyr) between bulge and disk.
\brem{d)} log(\mstotal) vs. metallicity contrast \dbdz\ (\mzmass\ -- \mdzmass) in \zsun\ between bulge and disk.}\label{bvsd}}
\end{picture}
\end{figure*}
\end{center}      
\subsection{Gas excitation mechanisms vs. bulge \dmb\ \label{sect3-b}}
The importance of various gas excitation mechanisms in the three \dmb\ intervals is examined in panel~\brem{i}, on the basis of BPT diagnostics within simulated 3\arcsec\ SDSS apertures (method \brem{a} in Sect.~\ref{BPT}). 
It can be seen that accretion-powered nuclear activity (reflected in Seyfert- and eventually also LINER-specific BPT line ratios) does not manifest itself uniformly along the bulge \dmb\ vs. \mbstar\ -- \sstar\ -- \mtmass\ sequence, but is confined to the \dmb\ interval \brem{iB}--\brem{iC}.
Whereas this is consistent with the previously reported scarcity of Seyfert activity in intermediate-to-low luminosity bulges that are commonly associated with PBs \cite[][for a review]{KormendyHo2013}, an insight from this panel is the association between physical and evolutionary properties of bulges with their activity status: lower \mbstar\ and \sstar\ \brem{iA} bulges fall almost exclusively in the locus of SF, whereas higher-mass ($\ga 10^{10}$ \msun), higher-\sstar\ ($\ga 10^{9}$ \msun\ kpc$^{-2}$) \brem{iC} bulges show in their majority Seyfert- and LINER-typical BPT ratios. 

Quantitatively, $\sim$93\% (38/41) of the bulges classified as Seyfert and LINER in our sample fall within the gray-shaded quadrant in panel~\brem{e} that depicts the area ($\log$\,\mbstar;$\log$\,\sstar) $\geq$ (10;9). 
Of the 76 LTG bulges in this locus of the diagram, 16 are classified as Composite, reflecting gas excitation by a mixture of accretion-powered and SF activity. 
As already mentioned in Sect.~\ref{BPT}, the proportion among different spectroscopic 
classes does not drastically change when area-weighted determinations within \rbulge\ (method~\brem{b}) are considered instead. These yield within the gray-shaded quadrant a proportion 23:11:22:20 between Seyfert, LINER, Composite and SF, lending further support to the conclusion that 
accretion-powered nuclear activity is primarily associated with old, high-\mbstar\ and high-\sstar\ bulges in the upper range of the \dmb\ sequence.
\subsection{Bulge-to-disk age and metallicity contrast \label{sect3-c}}
A comparative study of the evolutionary properties of bulges relative to those of their hosting disks may add further insights 
into the bulge growth process. 
In this regard, it is noteworthy that all bulges in our sample contain a non-negligible mass fraction $m_9$ (\%) of stars older than 9 Gyr, with average values of 21$\pm$0.10, 59$\pm$0.15 and 85$\pm$0.07 in the \dmb\ intervals \brem{iA}, \brem{iB} and \brem{iC}, respectively. The respective $m_9$ determinations for the disk (24$\pm$0.09, 36$\pm$0.08 and 49$\pm$0.11) suggest that the oldest bulges are hosted by the oldest disks, and vice versa. 

Indeed, a relation between the evolutionary properties of the disk (\rr$\geq$\rbulge) and the bulge is suggested from panels~\brem{a}\&\brem{b} of Fig.~\ref{bvsd}, that show the mass-weighted age and metallicity of the bulge and disk (large and small dots, respectively) as a function of total 
stellar mass \mstotal. It can be appreciated from linear fits to the data (solid lines) that LTG disks follow similar yet shallower trends with \mstotal\ as bulges, which is consistent with a synchronized evolution of the age and metallicity of the disk (\mdtmass\ and \mdzmass, respectively) with 
\mtmass\ and \mzmass. Our data yield the relations  
\mtmass\ = (1.21$\pm$0.06)$\cdot$\mdtmass\ -- (0.18$\pm$0.49)
and
\mzmass\ = (1.09$\pm$0.05)$\cdot$\mdzmass\ -- (0.05$\pm$0.03).

As better visible from panel~\brem{c}, approximately all bulges are on average older than the disk (the exception are the lowest mass LTGs whose bulge and disk are almost indistinguishable from one another with respect to their mean age and metallicity ).
Interestingly, the bulge-to-disk age contrast \dbdt\ (\mtmass\ -- \mdtmass) increases with \mstotal\ as
\dbdt\ (Gyr) = (1.00$\pm$0.12)$\cdot$log\,\mstotal\ -- (9.07$\pm$1.28). 
A similar, though weaker, trend can be appreciated from panel~\brem{d} for the bulge-to-disk metallicity contrast \dbdz\ (\mzmass\ -- \mdzmass) that scales with galaxy mass as \dbdz\ (\zsun) = (0.08$\pm$0.01)$\cdot$log\,\mstotal\ -- (0.75$\pm$0.12).
The mean \dbdt\ in the \dmb\ intervals \brem{iA}, \brem{iB} and \brem{iC} was determined to be 0.50$\pm$1.1, 1.65$\pm$0.7 and 2.07$\pm$0.6 Gyr. 
It is also interesting to note that the age contrast in \brem{iC} bulges shows a large spread between $\sim$0.5 and $\sim$3.6 Gyr, and that in up to $\sim$40\% of the bulges falling in the interval \brem{iA} it is slightly negative  ($\sim$--1 Gyr), hinting at a rise of SF activity in the bulge over the recent few Gyr. As expected from the evidence of Fig.~\ref{res}, \dbdt\ and \dbdz\ also show a positive correlation with \dmb.

Our conclusions above are in qualitative agreement with those by \citet{SB14} who find from 
a spectral modeling study of 62 nearly face-on spiral galaxies from CALIFA that bulges show, on average, higher luminosity-weighted 
ages and metallicities than disks. The latter study also shows that the metallicities of bulge and disk are correlated \citep[see also, e.g.,][]{MoHo06}, 
in agreement with the trend in Fig.~\ref{bvsd}b, whereas no correlation was found between the age of the two components \citep[see also,][]{SB16}.
Another insight from the study by \citet{SB14} is that the slope of the relation between central velocity dispersion 
and luminosity-weighted age and metallicity is similar for the central parts of bulges and at a galactocentric radius equivalent 
to $\sim$2.4 exponential disk scale lengths.

\section{Discussion \label{dis}}
In advance to the discussion below, it is worth recalling that this study combines, for the first time, three elements -- 
surface photometry, spectral modeling of IFS data and post-processing of population vectors with \RY\ -- 
toward a systematic investigation of the evolutionary properties of LTG bulges within a uniformly defined isophotal radius 
that encompasses nearly their total emission.
\subsection{The continuous rise of bulges out of galactic disks}
The essential insight from Sect.~\ref{res1} is that LTG bulges across $\sim$3 dex in \mstar\ and $>$1~dex in $\Sigma_{\star}$ 
form a \emph{continuous} sequence with regard to \dmb, \mtmass\ and \mzmass.
This argues against an age bimodality that would echo two distinct bulge assembly routes, 
one directing to old, monolithically formed CBs and the other one to PBs emerging purely through quasi-continuous SF 
in the centers of secularly evolving disks. The combined evidence suggests instead that bulges and disks evolve alongside 
in a concurrent process that leads to a continuum of physical and evolutionary properties being reflected in \dmb\ 
and exemplified by its three intervals (\brem{iA}--\brem{iC}) in Fig.~\ref{classes}. 
In high-mass (log(\mbstar)$>$10) LTG bulges (interval \brem{iC}), SF has shut off earlier 
than 9 Gyr ago, whereas in intermediate-mass ones (\brem{iB}) it was prolonged to a later cosmic epoch, 
with the least massive bulges (\brem{iA}) sustaining ongoing SF and being composed to $\sim$80\% of stars younger than 9 Gyr. 
Additionally, the fact that older disks in our sample host older bulges (judging from a comparison of $m_9$ ratios and panels \brem{a}\&\brem{c} of Fig.~\ref{bvsd}) with a within $\sim$0.2 dex equal metallicity (panel \brem{d}) points to an interwoven bulge--disk evolution,
as already suggested by some previous photometric or spectroscopic studies in the local universe and at higher redshifts \citep[e.g.,][see also Sect.~\ref{intro}]{Car07,vD13,Papovich15,SB14,SB16,Lan14,Tac17,Mar17}.

Our results are consistent with a picture where bulge growth in LTGs is driven by a superposition of quick-early and slow-secular processes, the relative importance of which increases with increasing galaxy mass.
Quick-early processes might comprise collapse of self-gravitating gas that leaves behind a massive 
nuclear cluster or a low-mass proto-bulge, both providing a steep gravitational pool that could promote
ensuing gas inflow and SF, eventually also the seed of a SMBH, in addition to inward migration and coalescence of massive 
($\ga 10^{8-9}$ \msun) SF clumps forming continuously out of violent disk instabilities 
\citep[e.g.,][see Sect.~\ref{intro}]{Bournaud07,Mandelker14,Mandelker17}. 
Inflowing inter-clump gas could add further fodder for the bulge buildup \citep[e.g.,][]{Hopkins2012,Zolotov15} 
during its dominant formation phase and beyond.
As for slow-secular contributions to the bulge growth, these may comprise, according to our current understanding, steady gas inflow from the disk sustaining a perpetual rejuvenation of the bulge through in situ SF, in concert with, for example, inward stellar migration and minor mergers of dwarf satellites (see Sect.~\ref{intro}). 
In terms of standard SFH parametrizations, the superposition of these processes might be empirically approximated by "delayed" exponentially 
declining SFR models \citep[e.g.,][]{CGP17} with an e-folding timescale and a delay of the SFR peak scaling inversely to the present-day \mstotal.
We note that the evidence from this study is qualitatively consistent with insights gained from Lick index studies by, for example, \citet{ThoDav06}, 
which indicate that the smallest bulges show the lowest light-weighted ages and a late enrichment in iron from SNe~Ia.

Clearly, a deeper understanding of the timescales and physical channels of bulge growth, including the possible regulatory role of AGN- and SF-driven feedback, is a fundamental yet formidable task that requires improved observational insights \citep[for instance, on the mass function of SF clumps in high-$z$ proto-disks; cf. e.g.,][]{Tamburello17,Cava17} and a detailed treatment of stellar-dynamical processes, such as stellar migration and its observational
imprints \citep[e.g.,][]{SB09,Roskar12,RL16}, the dissolution and regeneration of stellar bars \citep[e.g.,][]{Ber04}, and, more generally, of all 
intrinsic and environmental parameters that shape the dynamical coupling of stars, gas and dark matter (DM) in an evolving galaxy ecosystem.
For example, a process that might act toward steepening the density profile of a bulge, thereby possibly promoting 
nuclear starburst activity \citep{Loose82}, is adiabatic contraction of the stellar component in response to gas inflow.
This effect, which was proposed by \citet{P96b} as an explanation for the steeper stellar density profile of the underlying 
host of starbursting dwarfs, as compared to that of quiescent dwarf irregulars, might be relevant if accreted gas 
from the disk or directly from the cosmic web \citep[e.g.,][]{Dekel09,DekelBurkert14} 
contributes a significant ($\ga$30\%) fraction of the dynamical mass in the central parts of proto-LTGs.
This might be the case if baryons dominate in those radii 
\citep[see][for recent observational evidence for baryon dominated disks $\sim$10 Gyr ago]{Genzel17}, or the DM halo has an extended core of nearly constant density \citep[i.e., a Burkert 1995 profile, see also][]{SalucciBurkert00}.

\subsubsection{Star formation rate vs. stellar mass growth rate}
A subject of considerable interest is migration and its possible role on the mass growth rate (\mgr) of LTG bulges. A subtle yet important point is that the \mgr\ is not necessarily equal to the SFR, since inward (outward) migration across \rbulge\ may lead to a higher (lower) \mgr\ than the value implied by in situ SF, after correction for the stellar mass fraction returned into the interstellar medium in the course of stellar evolution.
The case of \mgr$<${\sc sfr} was discussed in \citet[][see also Papaderos \& \"Ostlin 2012]{P02} who conjectured that outward diffusion of newly formed stars in the starbursting cores of young dwarf galaxies may be an important driver of the buildup of their redder, more extended LSB stellar host: since the high-mass end of stellar populations with a lifetime shorter than a migration timescale \mit\ will be depopulated before reaching the LSB periphery, a stellar mass filtering effect will naturally lead to a radial (core-envelope) age (and color) gradient. 
A consequence of this is the overestimation of age for the LSB host, if its colors (or spectrum) are interpreted on the usual assumption of in situ SF according to a fully populated, spatially invariant initial mass function. The inverse situation (\mgr$>${\sc sfr}) might occur in the case of inward migration of ex situ formed SF clumps and stars from the disk, reaching the bulge with a delay \mit\ (presumably, greater than several $10^8$ yr) after their formation. 
Since this process imposes a minimum age \mit\ for stellar populations arriving at \rbulge, inward migration into old, high-\mbstar\ and --\sstar bulges could under certain circumstances act toward decreasing \dmb, $m_9$ and the bulge-to-disk age contrast \dbdt. Following these considerations, the absence of stellar populations younger than 9 Gyr in \brem{iC} bulges does not necessarily rule out that their buildup has been continuing until (9-\mit) Gyr ago. Conversely, the presence of stars of age \mit\ in low-mass bulges does not imply that in situ SF in these entities has been ongoing already \mit\ Gyr ago. Summarizing, one effect that might be plausibly expected from inward migration is the aging (rejuvenation) of young (old) bulges by $\simeq \xi\cdot \langle$\mit$\rangle$, whereby $\xi$ denotes the mass fraction of inwardly migrated stars or SF clumps, and $\langle$\mit$\rangle$ is a mass-weighted average migration time for stars arriving in the bulge from different radii of the disk.

The sketchy remarks above are merely meant to outline one principle effect that inward migration might have on bulge age and SFH determinations, 
and to highlight the need for a better observational and theoretical understanding of this process. This task appears to be of considerable interest in the light of, for example, the recent conclusion by \citet{Mar17} that $\sim$1/2 of the stellar mass formed in the disks of two-component galaxies 
at $1 < z < 3$ must have transferred inward, promoting the growth of the bulge.
\subsubsection{Accelerated bulge growth in proportion to galaxy mass \label{bulgegrowth}}
From our previous remarks it follows that an enhanced \mgr\ in the bulge could drive a non-homologous growth of radial \tsstar\ and $\mu$ profiles in LTGs: starting from an almost bulgeless  exponential proto-disk with a S\'ersic index $\eta\simeq1$, an inwardly increasing \mgr\ would act toward accelerating the buildup of the bulge and in turn lead to a steepening of the galaxy profile ($\eta > 1$) and gradual increase of the \mbstar/\mstotal\ ratio. This is consistent with Fig.~\ref{res}c that shows that high-\dmb\ (\brem{iC}) bulges hosted by massive 
(log\,\mstar$\simeq$11) LTGs contain twice the \mstotal\ fraction than low-\dmb\ (\brem{iA}) bulges ($\sim$40\% and $\sim$20\%, respectively). The mass increase and compaction of the bulge to a high \tsstar\ might also support the central confinement and collapse of inflowing gas within its steepening gravitational potential, enhance chemical self-enrichment and therefore the gas cooling efficiency \citep[cf., e.g.,][]{BH89}, which could in turn accelerate the SF cycle and the growth of \mbstar\ in a manner that is proportional to the local \tsstar. This process might continue  until SF starts being gradually extinguished, once the bulge has attained a ($\log$\,\,\mbstar;$\log$\,\,\sstar) $\geq$ (10;9) 
and accretion-powered nuclear activity becomes important (cf. Sect.~\ref{agn}).
The rise of bulges out of disks might therefore bear some analogy to (sub)galactic downsizing, in the sense that the assembly timescale of stellar systems scales inversely with the mass and initial density of the baryonic condensations galaxies emerge from, in qualitative resemblance to the staged galaxy formation scenario by \cite{Noeske07a}. 
 
Besides the trend for increasing \dmb\ and \mtmass\ with increasing \mbstar\ and \mstotal\ (Fig.~\ref{res}a-c), further support 
to the notion of an accelerated bulge growth in proportion to the LTG mass comes from Fig.~\ref{bvsd}c that shows the relation 
between bulge-to-disk age contrast \dbdt\ as a function of \mstotal: would bulges form purely monolithically, prior to and independently of disks, then one would expect a roughly constant ($>$0) \dbdt\ across LTG mass If, on the other hand, bulges were assembling strictly in lockstep with secularly evolving disks, thus sharing the same SFH and essentially being rearranged disk material, then their \dbdt\ would also be independent on \mstotal\ and in the range between $\approx$0 Gyr and $\langle$\mit$\rangle$.
However, the documented trend for an increasing bulge-to-disk age contrast with increasing \mstotal\ suggests that higher-mass bulges in higher-mass LTGs have assembled earlier and quicklier, and vice versa. We note that the relation between \dbdt\ and \mstotal\ (or, equivalently, \dmb; cf. Sect.~\ref{res1}) is unlikely to be driven by the earlier cessation of SF in higher-mass disks, since this would tend to diminish \dbdt, contrary to the evidence from Fig.~\ref{bvsd}c.
Quite importantly, the absence of a bimodality or even discontinuity in the bulge vs. disk age and metallicity contrast over $\sim$2.5 dex in \mstotal\ points to an interwoven 
yet asynchronous growth of these components, with bulges assembling out of disks on a timescale that scales inversely with \mstotal.
In this sense, our results support neither the monolithic nor the purely secular bulge formation scenario, but point instead to a picture of sub-galactic downsizing 
where bulges form faster in more massive baryonic entities, and vice versa.
The non-homologous radial growth of \tsstar\ implied by this bulge growth process is also consistent with the observed trend 
between the \mbstar/\mstotal\ ratio and \mstotal\ (Fig.~\ref{res}c), 
as well as with a trend for an increasing S\'ersic index $\eta$ with \mstotal\ (or $M_r$) for present-day LTGs.

The three LTGs in Fig.~\ref{classes} may be regarded as morphological snapshots in a chronological sequence of increasing \mbstar\ and \sstar\ in tandem with \dmb. Despite the tentative subdivision of LTG bulges into three intervals (\brem{iA}--\brem{iC}), for the sake of illustration, it is important to bear in mind that the here documented continuity in the bulge properties themselves and their difference (\dbdt\ and \dbdz) to their hosting disks consistently disfavors a bulge segregation into evolutionary distinct classes, in particular between CBs and PBs.
One may frame the insights from this study in a simple scenario where high-mass bulges (\dmb\ interval \brem{iC}) represent the endpoint of an evolutionary pathway across an increasing \dmb\ (\brem{iA}$\rightarrow$\brem{iB}$\rightarrow$\brem{iC}) whose duration depends on the intrinsic and environmental properties of the halo in which LTGs form: higher-mass systems form preferentially in higher-density environments, where efficient gas accretion from the cosmic web \citep[e.g.,][]{Dekel09} facilitates rapid mass growth and complete the dominant phase of their bulge assembly earlier than 9 Gyr ago, while eventually maintaining a significant \mgr\ in the ensuing Gyrs of their passive photometric evolution through inward stellar migration and minor dry mergers.
\begin{figure}
\includegraphics[width=1\linewidth]{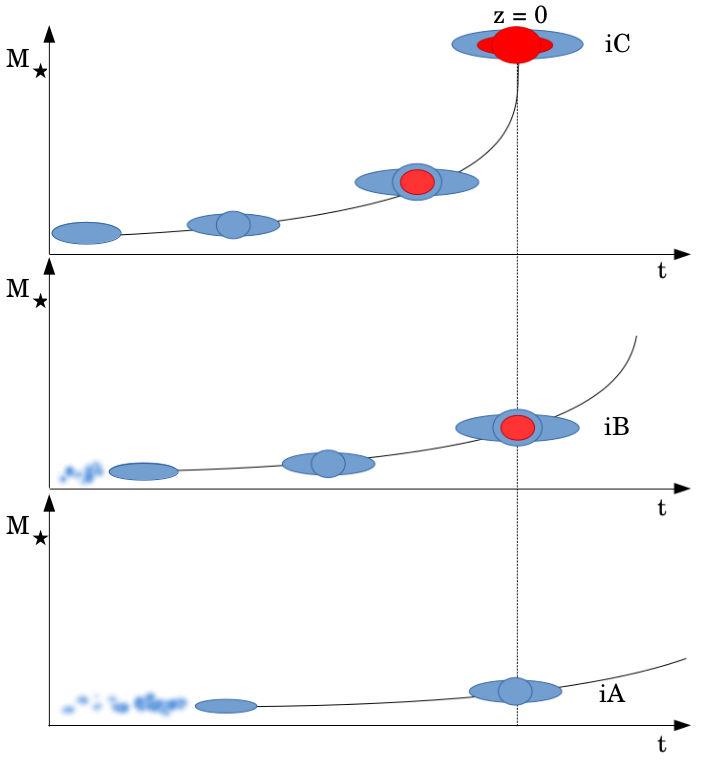}
\caption[Schematic illustration of the scenario envisaged here]{Schematic illustration of the scenario proposed in Sect.~\ref{bulgegrowth}: the three LTGs in Fig.~\ref{classes} may be regarded as morphological snapshots in a sequence of increasing bulge stellar mass \mbstar, age \mtmass, surface density
\sstar\ and \dmb. The progenitors of high-mass LTGs (e.g., \object{NGC 0776}, i.e., systems in the \dmb\ interval \brem{iC}; $\geq$-0.5 mag) form in more massive halos and experience the dominant phase of their bulge assembly earlier than 9 Gyr ago (vertical line). Conversely, the precursors of low-\dmb\ ($\leq$-1.5 mag; i.e., \brem{iA}) LTGs, such as \object{IC 0776}, assemble at a calm pace within lower-mass halos and undergo an overall retarded structural and chemical evolution toward intermediate-\dmb\ LTGs (\brem{iB}), such as \object{NGC 0001}.
Once the bulge has grown to (log\,\,\mbstar;\,log\,\,\sstar) $\geq$ (10;9), AGN-driven feedback initiates a gradual cessation 
of SF (depicted in red), while the bulge mass may continue increasing through inward stellar migration from the disk and multiple minor dry mergers. 
}\label{sketch}
\end{figure}
Conversely,  LTGs born out of lower-mass halos in less 'privileged' regions of the cosmic web (filaments and their surroundings) might have experienced a delayed buildup as a result of comparatively inefficient gas accretion onto their shallow gravitational pools.
The low 'metabolism' (aka sSFR) of these latecomers in the \brem{iA} interval of \dmb\ is imprinted on their lower age, metallicity, stellar surface density and \mbstar/\mstotal\ ratio. Moreover, the youth of these systems is not at odds with the presence of a minor fraction ($\sim$20\%) of old ($>$9 Gyr) stars (cf. Sect.~\ref{sect3-c}) that could originate from low-level SF and minor mergers with ancient galaxy building blocks, such as, for example dwarf spheroidals formed prior to the reionization epoch \citep[e.g.,][]{GG04}. 

It is interesting to note that results from this study are qualitatively in line with conclusions drawn from abundance matching studies 
of higher-$z$ galaxies \citep[e.g.,][]{vD13,Patel13,Papovich15}. For instance, \citet{Papovich15} have investigated 
how the progenitors of present-day LTGs with the mass of the MW and M31 ($5 \times 10^{10}$ \msun\ and $10^{11}$ \msun, respectively) 
may have grown from $z=3$ to 0.5. 
They report that both started their evolution as star-forming disks, and after an IR-luminous phase, they evolved in the redder and more quiescent galaxies of today. These authors further conclude that "the progenitors of MW-mass galaxies reached each evolutionary stage at later times (lower redshifts) and with stellar masses that are a factor of two to three lower than the progenitors of the M31-mass galaxies.". 
Furthermore, these studies report a smooth increase of S\'ersic $\eta$ with $z$, with higher-mass galaxies 
reaching a higher $\eta$ earlier than lower-mass galaxies, in agreement with the hypothesis of an accelerated bulge growth 
in proportion to galaxy mass suggested above.
\subsection{Accretion-powered nuclear activity along the bulge sequence \label{agn}}
The occurrence of accretion-powered nuclear activity in galactic bulges has been the subject of intense investigation over the past decades 
\citep[the reader is referred to][for a recent review]{KormendyHo2013} showing that manifestations of weak Seyfert activity are comparatively 
rare in disk-dominated LTGs that are commonly classified as PBs. Given the kinematical evidence for these bulges also containing  
a SMBH, just like CBs, the observed scarcity of AGN activity has been generally ascribed to an inefficient (sub-Eddington) and episodic matter accretion \citep[e.g.,][]{Kor11} or to the premature termination of SMBH growth, possibly due to the dissolution of short inner bars \citep{Du17}.

This picture is echoed in Fig.~\ref{res}i, showing that 31 of the 33 lower-mass \brem{iA} bulges in our sample fall in the SF locus of the BPT diagram, 
with only two of them being spectroscopically classified as Composite. Assuming that the $M_{\bullet}$/\mstotal\ ratio of $\approx 10^{-3}$ for galaxies with 
log(\mstotal)$\geq$10 \citep[][]{KormendyHo2013} applies to lower masses, 
the $M_{\bullet}$ of $\sim 10^{6-7}$ \msun\ expected for \brem{iA} bulges in our sample
would translate to maximum Eddington rates of 0.02-0.2 \msun/yr, assuming a radiative efficiency of 0.1. 
Even though weak and smeared by the PSF of CALIFA IFS data, an AGN point source powered by these accretion rates 
should probably be detectable above the local background of these least massive bulges, given that their surface brightness 
is by 2--3 SDSS $r$ mag fainter than that of high-mass (\brem{iC}) bulges. 
Moreover, the omnipresence of SF in these bulges, implying a cold gas reservoir of sufficient density, 
suggests that the absence of Seyfert signatures is not due to a high Ly$_{\rm c}$ continuum photon escape 
fraction (contrary to the case of many massive ETGs) or their dilution by circumnuclear SF (cf. discussion in Sect.~\ref{BPT}).
Moreover, an obscured AGN does neither offer a convincing explanation, given the relatively strong nebular emission 
(EW(\ha)=26.3 $\AA$, on average), moderate-to-low ($A_V\leq 0.24\pm0.12$ mag) intrinsic extinction and nearly face-on geometry of these low-\mbstar\ LTGs.
The evidence from Fig.~\ref{res}i is certainly compatible with the picture of inefficient or sporadic gas accretion onto an intermediate-$M_{\bullet}$
SMBH, as suggested from previous work. A conceivable alternative, on the other hand, might be an asynchronous growth of SMBHs 
relative to their galaxy hosts with the $M_{\bullet}$/\mbstar\ ratio, or the Eddington ratio and the SMBH spin parameter 
being non-linearly coupled to \mstotal.

Coming to massive LTG bulges with a \dmb\ above --1.5 mag (\brem{iB} and \brem{iC}), we recall that the gray-shaded quadrant in Fig.~\ref{res}e 
contains 93\% (38/41) of all Seyfert and LINER in our sample. This suggests that the deep gravitational potential of bulges in the range 
log(\mbstar;\sstar)$\geq$(10;9) is linked, perhaps causally, to an efficient SMBH feeding.
The regulatory role of the associated AGN phenomenon on SF quenching in these old, high-mass bulges remains a subject of intense investigation
\citep[e.g.,][and references therein]{Cattaneo2009,KormendyHo2013,Har17} both in the local universe and at intermediate $z$.
For instance, \citet{Lan14} study 6764 galaxies with \mstotal$>10^{10}$ \msun\ at $0.5 < z < 2.5$ by performing bulge-to-disk decomposition. 
Their findings suggest that bulges and SMBHs grow hand in hand through merging and/or disk instabilities, with AGN-feedback being 
the main driver for the shut-off of SF. \cite{Tac17} conjecture that SF quenching mechanisms must be internal to the galaxies and closely associated with bulge growth, and that, at \mstotal\ $> 10^{11}$ \msun, SF quenching gradually progresses in an inside-out manner.
This conclusion appears to be consistent with our results: 16 of the 76 LTGs in the gray-shaded area 
of Fig.~\ref{res}e are classified as Composites, which points to the co-existence of SF with AGN-driven feedback over a time span of $\ga$2 Gyr 
(cf. panel \brem{b} with panel \brem{f}), therefore against an abrupt SF quenching by accretion-powered nuclear activity.

\section{Summary and conclusions \label{conc}}
This study combines for the first time three techniques -- surface photometry of SDSS data, 2D analysis and spectral modeling of IFS data and post-processing of the spectral synthesis output with the code \RY\ -- toward  a systematic investigation of the physical and evolutionary properties (e.g., \mstar, \tstar, \tsstar, \zstar, $M_{r}$, \rbulge) of galaxy bulges within a uniformly defined isophotal radius that encompasses almost their total emission. 
Our sample is composed of 135 non-interacting, nearly face-on LTGs from the CALIFA survey 
that densely cover a wide range in bulge stellar mass (10$^{8.3}$ -- 10$^{11.3}$ \msun) and luminosity ($-20\la M_r \la -14$).

A central element of our analysis revolves around \dmb\ (mag), a quantity that can be inferred from \RY\ and which allows to evaluate the contribution of stellar populations older than 9 Gyr to the mean $r$-band surface brightness of galaxy bulges. We show that this distance- and nominally extinction-independent quantity, first introduced and systematically investigated for a large LTG sample here, offers a simple yet valuable semi-empirical proxy to the bulge assembly history, and eventually a promising bulge classification diagnostic.

The main results of this study may be summarized as follows:
\begin{itemize}
\item[i.] LTG bulges span a range in \dmb\ between $\sim$--4 mag and $\sim$0 mag, translating, respectively, into a contribution to the bulge mean surface brightness between $\sim$3\% and $\sim$100\% by stellar populations older than 9 Gyr.
Quite importantly, \dmb\ follows a tight correlation with the present-day stellar mass \mbstar, stellar surface density \sstar, mass-weighted age \mtmass\ and metallicity \mzmass\ of LTG bulges over $\sim$3 dex in \mbstar\ and $>$1 dex in \sstar: The highest-\dmb\ ($\sim$0 mag) bulges are the oldest, most massive, dense and chemically enriched, and vice versa. 
\item[ii.] On the basis of \dmb, we tentatively subdivide LTG bulges in three characteristic intervals: low-mass (log\,\mbstar$\la$9.7) bulges with a \dmb $\leq$ --1.5 mag (interval \brem{iA}) are hosted by low-mass (log\,\mstotal$\la$10.5) disk-dominated LTGs, are comparatively young (\mtmass$\simeq$6.8 Gyr) and show ongoing SF with an average EW(\ha)$\simeq$25 \AA. 
Conversely, high-mass bulges (\dmb\ $\geq$ --0.5 mag; interval \brem{iC}) reside in high-mass (log\,\mstar $\ga$ 11) LTGs, have a mean age of $\simeq$ 11.5 Gyr and are characterized by a high stellar mass surface density (\sstar$\geq 10^9$ \msun\,kpc$^{-2}$) and weak nebular emission ($\langle$EW(\ha)$\rangle\simeq$5 \AA. 
As for bulges in the intermediate range of \dmb\ (interval \brem{iB}), they also display intermediate values in \mbstar, \sstar, \mtmass, \mdzmass, absolute $r$-band magnitude and EW(\ha).
\item[iii.] Whereas more massive bulges are hosted by more massive LTGs, the bulge-to-total mass ratio \mbstar/\mstotal\ increases with LTG mass 
\mstotal, being on average twice as large in high-\dmb\ \brem{iC} bulges than in low-\dmb\ \brem{iA} bulges ($\sim$0.4 and $\sim$0.2, respectively).
\item[iv.] The age and metallicity of LTG bulges is moderately correlated with the age and metallicity of their parent disks.
However, the bulge-to-disk age and metallicity contrast increases with \mstotal\ and \dmb: 
whereas the bulge and the disk are of similar age and metallicity in low-mass LTGs, high-mass bulges are, on average, by 
$\sim$2 Gyr older and by 0.9 dex more metal-rich than the disk. 

\item[vi.] An analysis of BPT diagnostics indicates that SF is the dominant gas excitation mechanism in lower-mass bulges. 
Accretion-powered nuclear activity, manifesting itself in Seyfert- and eventually LINER-specific BPT line ratios 
is almost exclusively confined to higher-mass ($\ga 10^{10}$ \msun), higher-\sstar\ ($\ga 10^{9}$ \msun\ kpc$^{-2}$) bulges.\\
\end{itemize}

The essential insight from this study is that LTG bulges across $\sim$3 dex in \mstar\ and $>$1~dex in $\Sigma_{\star}$ 
form a \emph{continuous} sequence with regard to \dmb, \mtmass\ and \mzmass.
This argues against an age bimodality that would reflect two distinct bulge assembly routes, 
one directing to old, monolithically formed CBs and the other one to PBs emerging purely through quasi-continuous SF 
in the centers of secularly evolving disks. The combined evidence suggests instead that bulges and disks evolve alongside 
in a concurrent process that leads to a continuum of physical and evolutionary properties being closely reflected in \dmb.
Our results are consistent with a picture where bulge growth in LTGs is driven by a superposition of 
quick-early and slow-secular processes, the relative importance of which is increasing with increasing galaxy mass.

Furthermore, the trend for an increasing \mbstar/\mstotal\ ratio and bulge-to-disk age contrast with increasing \mstotal\ (or, equivalently, \dmb), and the absence of a bimodality or even discontinuity in this relation over three dex in bulge mass, lend support to the interpretation of an interwoven yet asynchronous growth of bulge and disk, with the former assembling out of the latter on a timescale that is inversely related to \mstotal.
This process is expected to lead to a non-homologous radial growth of \tsstar\ in LTGs and a trend for an increasing S\'ersic index $\eta$ 
with increasing \mstotal.

This framework points against a fundamental evolutionary dichotomy between CBs and PBs, instead unifying LTG bulges into a continuous chronological sequence of increasing \mbstar\ and \sstar\, and inviting to a further exploration of the mechanisms behind the non-homologous growth of stellar mass in galaxies and the rise of bulges out of galactic disks.

\begin{acknowledgements}
We thank the anonymous referee for valuable suggestions and comments.
This work was supported by Funda\c{c}\~{a}o para a Ci\^{e}ncia e a Tecnologia (FCT) through national funds and by FEDER through COMPETE by the grants UID/FIS/04434/2013 \& POCI-01-0145-FEDER-007672 and PTDC/FIS-AST/3214/2012 \& FCOMP-01-0124-FEDER-029170. We acknowledge supported by European Community Programme ([FP7/2007-2013]) under grant agreement No. PIRSES-GA-2013-612701 (SELGIFS). 
IB was supported by the fellowship PD/BD/52707/2014 funded by FCT (Portugal) and POPH/FSE (EC) and by the fellowship CAUP-07/2014-BI in the context of the FCT project
PTDC/FIS-AST/3214/2012 \& FCOMP-01-0124-FEDER-029170. PP was supported by FCT through Investigador FCT contract IF/01220/2013/CP1191/CT0002.
Linear fits were computed with the code OK, which was developed by Dr. Hans-Hermann Loose ($\dagger$ August 5th, 1993). 
We thank Dr. Jean Michel Gomes for useful comments. This study uses data provided by the Calar Alto Legacy Integral Field Area (CALIFA) survey (http://califa.caha.es), 
funded by the Spanish Ministry of Science under grant ICTS-2009-10, and the Centro Astron\'omico Hispano-Alem\'an.
It is based on observations collected at the Centro Astron\'omico Hispano Alem\'an (CAHA) at Calar Alto, operated jointly 
by the Max-Planck-Institut f\"ur Astronomie and the Instituto de Astrof\'isica de Andalucía (CSIC).
This research has made use of the NASA/IPAC Extragalactic Database (NED) which is operated by the Jet Propulsion Laboratory, 
California Institute of Technology, under contract with the National Aeronautics and Space Administration.
\end{acknowledgements}



\begin{appendix}

\section{An empirical assessment of disk contamination and PSF convolution effects on the obtained trends\label{A-check1}}
Even though all but one (\dmb) of the quantities involved in this study are weighted by \mstar, therefore relatively insensitive to the luminosity 
contribution by the star-forming disk, it is worthwhile to check whether the latter has an appreciable impact on \mtmass\ and \mzmass\ determinations within the bulge radius \rbulge. In particular, since the B/T mass ratio increases by a factor of $\sim$2 across the range in \mstotal\ covered by the analyzed LTG sample ($10^9$ \msun\ -- $10^{11.5}$ \msun), it cannot be excluded that the degree of disk contamination scales inversely with \mstotal, therefore primarily affecting low-B/T (i.e., low-\dmb\ \brem{iA}) galaxies and leading to an artificial steepening of the log(\mbstar) vs. \mtmass\ relation in Fig.~\ref{res}b.

For this purpose, a series of tests was made to study the dependence of \mtmass\ and \mzmass\ on the isophotal aperture considered 
by repeating the analysis for the three innermost \brem{isan}, as well as within the central portion of the bulge where contamination 
by the disk should be minimal. For the latter tests we adopted a radius of 3\farcs3 (hereafter $R_{\rm 3.3}$), which, given the typical full width at half maximum (FWHM) of CALIFA DR3 data ($\simeq$2\farcs6), encompasses the total emission of a point-like source (e.g., a compact bulge). 

As apparent from Fig.~\ref{A-R3.3}, mass-weighted age and metallicity determinations within $R_{\rm 3.3}$ do not appreciably differ 
from those within \rbulge, which suggests that the trends in Figs.~\ref{res}\&\ref{bvsd} are not notably affected by a \mstotal-dependent 
light contamination of the bulge by the surrounding star-forming disk. 

\setlength{\unitlength}{1mm}
\begin{center}
\begin{figure*}[b]
\begin{picture}(180,100)
\put(5,50){\includegraphics[height=5cm]{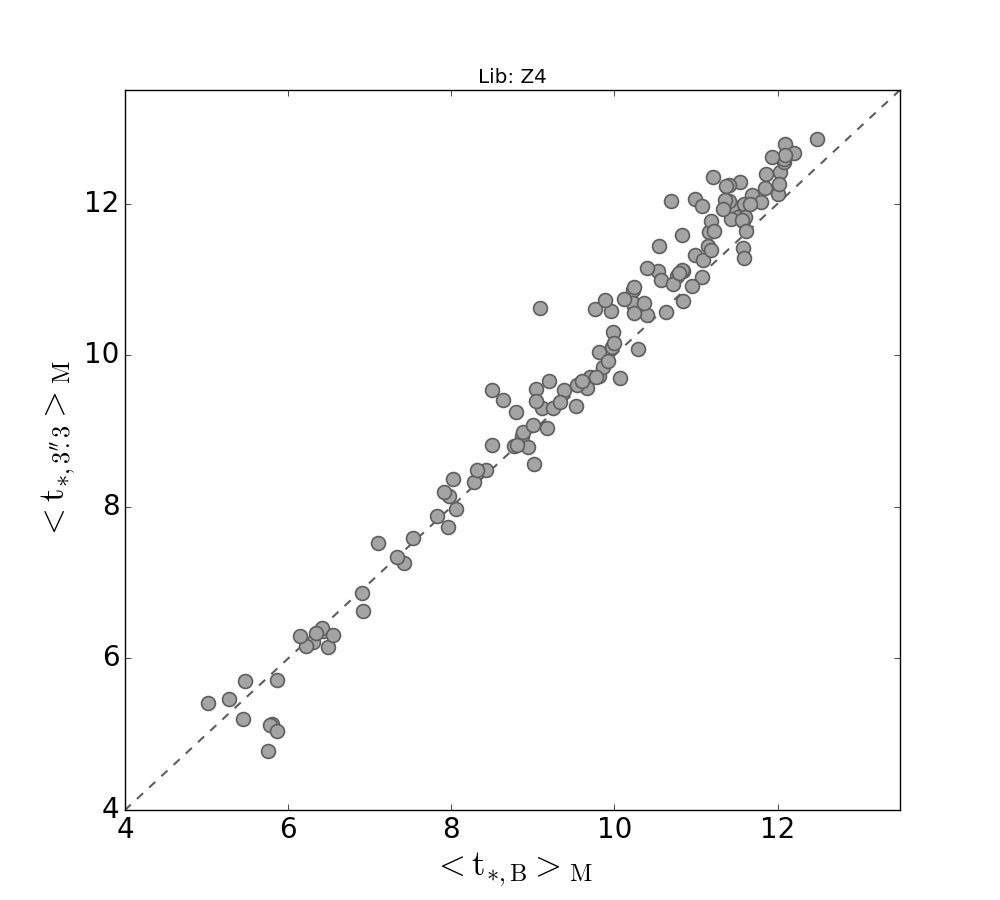}}
\put(60,50){\includegraphics[height=5cm]{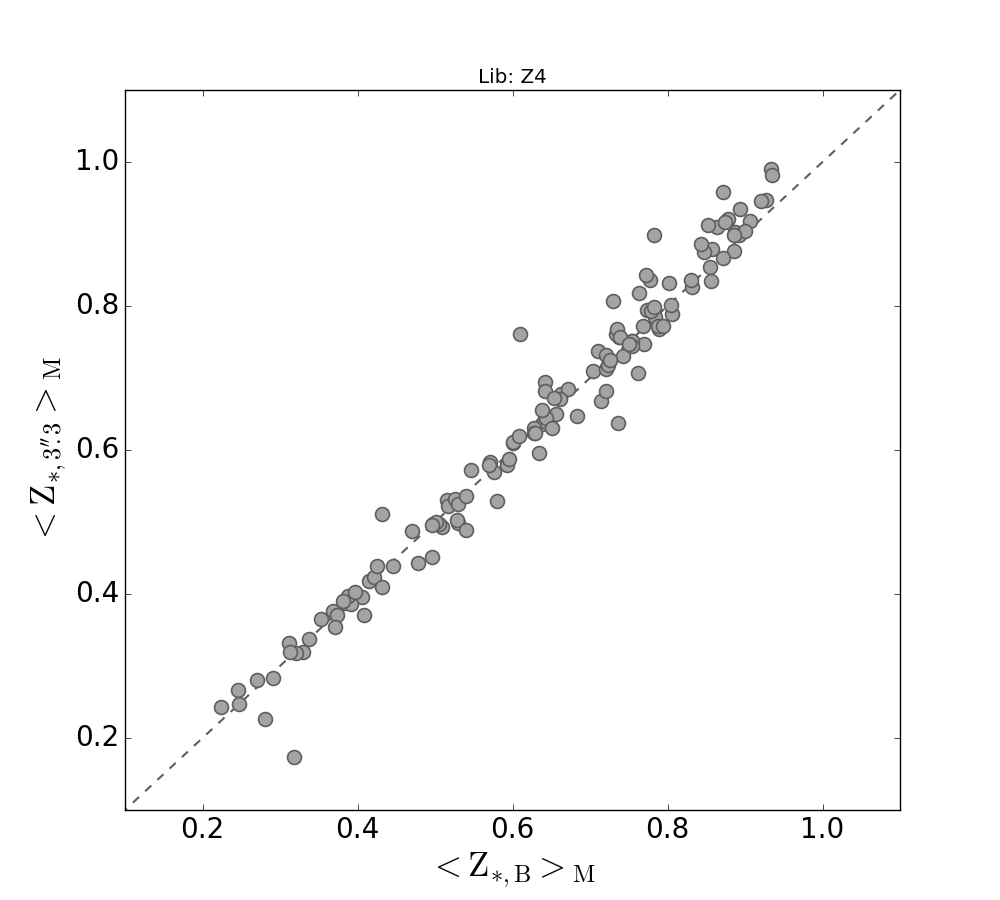}}
\put(5,0){\includegraphics[height=5cm]{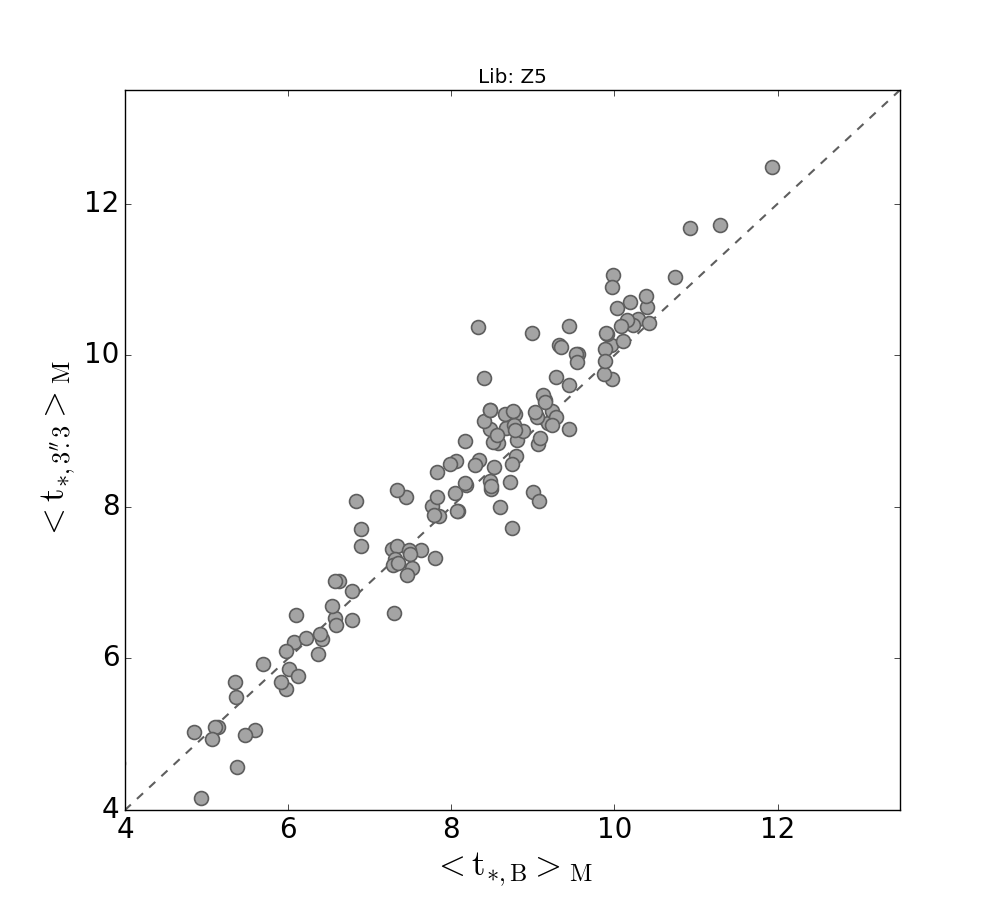}}
\put(60,0){\includegraphics[height=5cm]{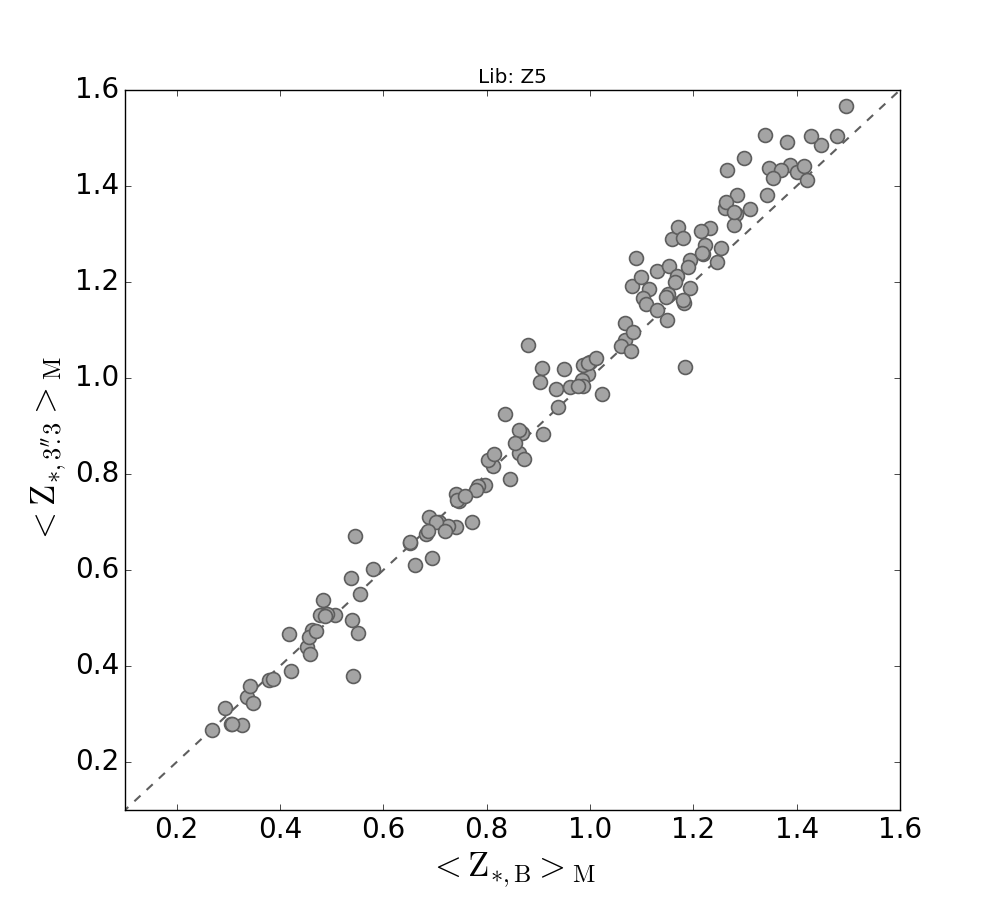}}
\PutLabel{15}{90}{\hvss a}
\PutLabel{70}{90}{\hvss b}
\PutLabel{15}{38}{\hvss c}
\PutLabel{70}{38}{\hvss d}
\PutWin{120}{72}{6cm}{
\caption{Comparison of the mean mass-weighted stellar age \mtmass\ (Gyr) and metallicity \mzmass\ (\zsun) within 
\rbulge\ (cf. Sect.~\ref{SpectralModeling}) with the values obtained within a radius of 3\farcs3 for the LTG bulges in our sample
($\langle t_{\star,\textrm{3.3}} \rangle_{{\cal M}}$ and $\langle Z_{\star,\textrm{3.3}} \rangle_{{\cal M}}$, respectively). 
Dashed diagonal lines indicate equality between both plotted quantities
for spectral models based on the Z4 and Z5 SSP library (upper and lower panels, respectively).
It can be seen that the scatter of points relative to the equality lines is relatively small
(standard deviation of 0.56 Gyr and 0.33 dex for spectral fits with the Z4 SSP library and 
0.63 Gyr and 0.60 dex for spectral fits with the Z5 SSP library) and do not show 
a significant dependence on age and metallicity.}}
\end{picture}
\label{A-R3.3}
\end{figure*}
\end{center}      

The good agreement between \mtmass\ and \mzmass\ determinations within $R_{\rm 3.3}$ and \rbulge\ is probably not due to point spread function (PSF) smearing effects, but presumably the result of weak radial gradients of mass-weighted quantities within LTG bulges. 
Indeed, the bulge radius in the majority ($\sim$95\%) of the analyzed LTGs is by a factor $\sim$2--4.6 larger than the FWHM of CALIFA IFS data (vertical dashed line in panel a of Fig.~\ref{A-FWHM}). Also, panels b\&c show that \rbulge\ is typically $\sim$2--3 times larger than the effective radius \reff\ and contains in all cases at least 80\% of the total emission of the bulge. 
\setlength{\unitlength}{1mm}
\begin{center}
\begin{figure*}[!ht] 
\begin{picture}(180,50)
\put(5,0){\includegraphics[height=5cm]{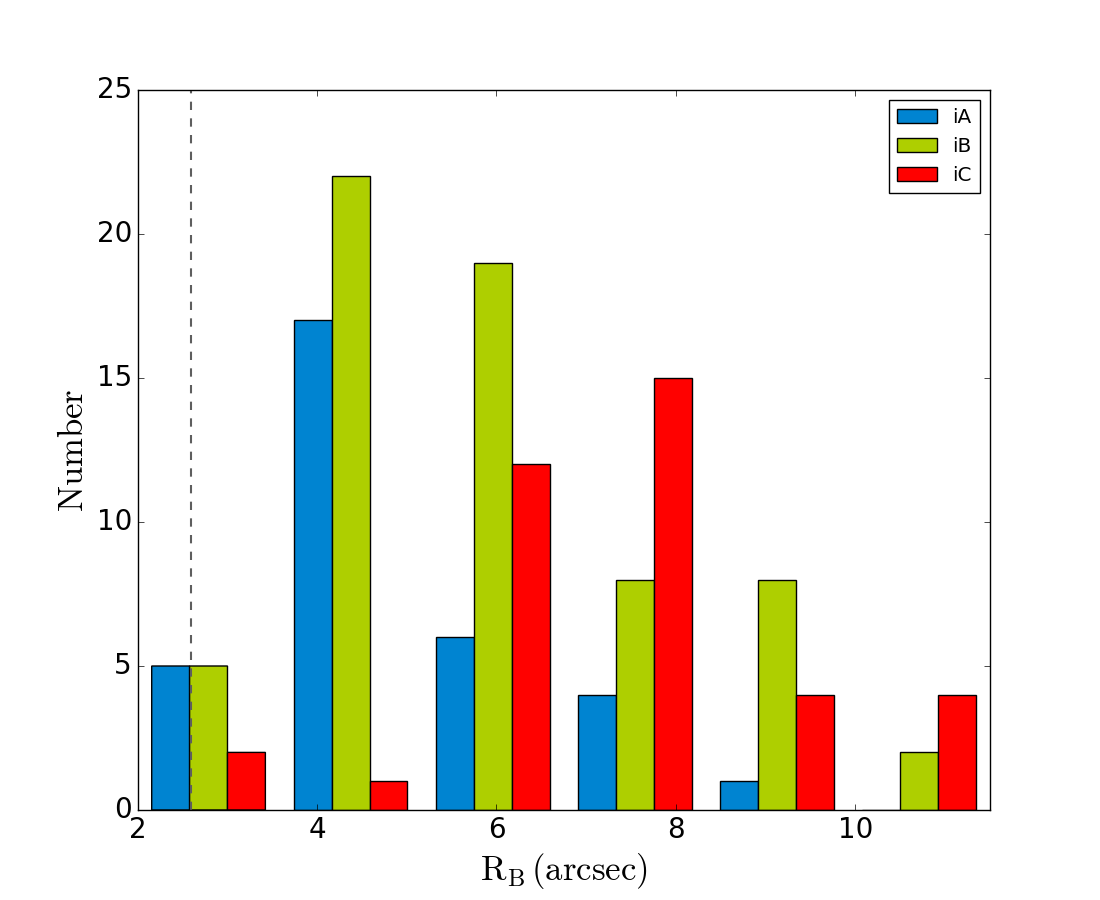}}
\put(60,0){\includegraphics[height=5cm]{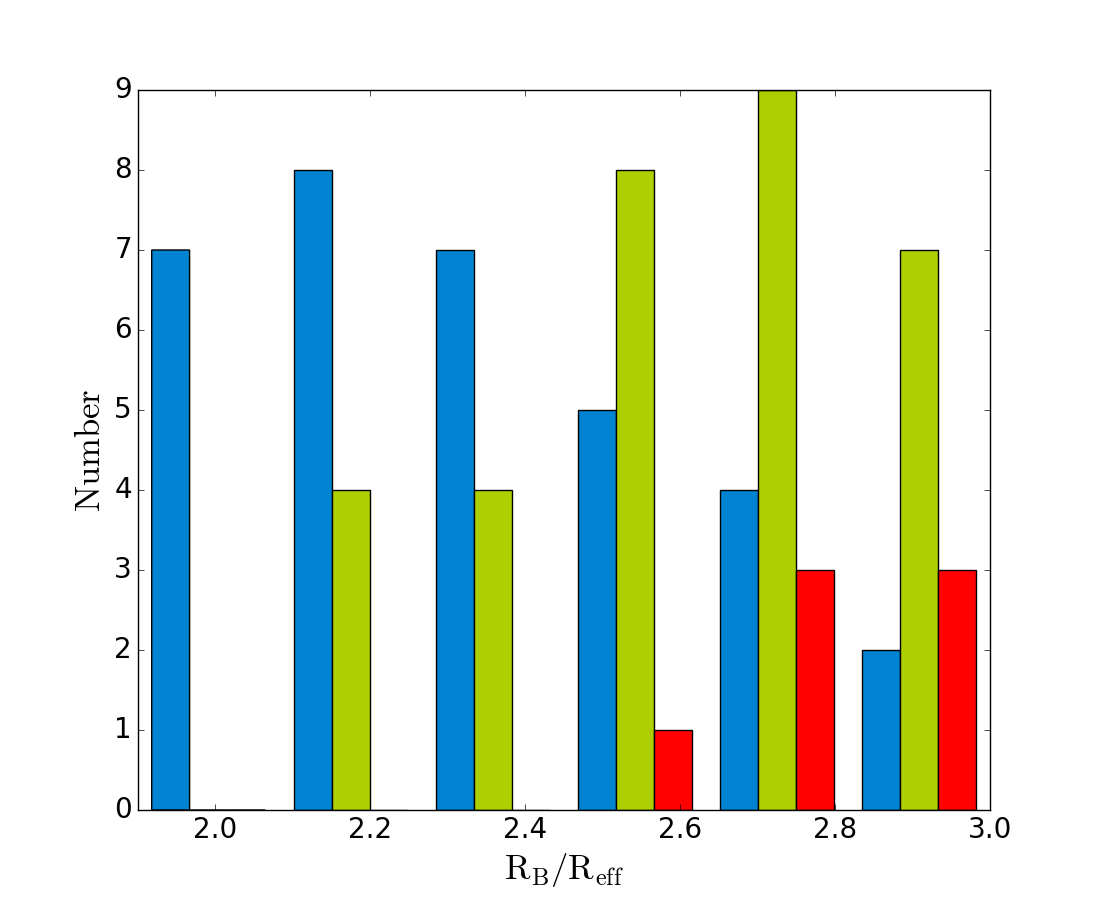}}
\put(120,0){\includegraphics[height=5cm]{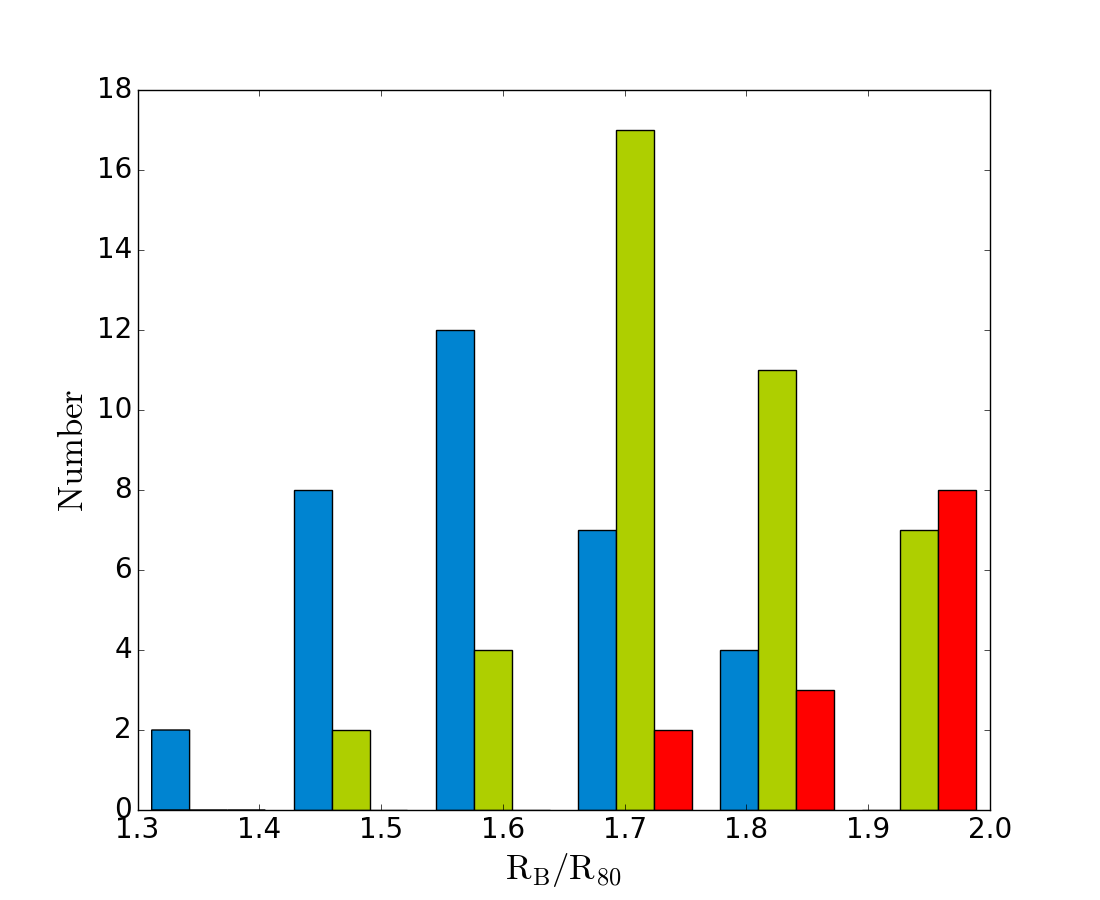}}
\PutLabel{18}{40}{\hvss a}
\PutLabel{70}{40}{\hvss b}
\PutLabel{130}{40}{\hvss c}
\end{picture}
\caption{Histogram distributions of the bulge isophotal radius \rbulge\ (\arcsec) of the LTGs in our sample (panel \brem{a}), and 
its normalized values to \reff\ and the radius enclosing 80\% of its total bulge luminosity (panels (panel \brem{b} and \brem{c}, respectively).
The dashed vertical line in panel \brem{a} marks the FWHM of CALIFA DR3 data (2\farcs6). The color coding is identical to that in 
Fig.~\ref{res} and is meant to illustrate the variation of the bulge extent among the three tentatively defined \dmb\
intervals (cf. Sect.~\ref{res1}).} 
\label{A-FWHM}
\end{figure*}
\end{center}      

\section{Dependence of the inferred trends on the metallicity range covered by the SSP library \label{ap:z}}

As pointed out in Sect.~\ref{SpectralModeling}, in our study preference was given to determinations
that are based on spectral modeling with the Z4 SSP library with the goal of improving age determinations 
at the expense of a possible saturation of the metallicity at $\leq$ $Z_{\odot}$. 
Nevertheless, a parallel spectral modeling analysis of the LTG sample was carried out using the Z5 SSP library, which additionally comprises
SSPs for a stellar metallicity of 1.5$\cdot$\zsun, in order to evaluate the influence of the AMD on the trends discussed in Sect.~\ref{res1}.

\setlength{\unitlength}{1pt}
\begin{center}
\begin{figure*}[b]
\begin{picture}(180.4,310.0)
\put(0,0){\includegraphics[width=1\linewidth]{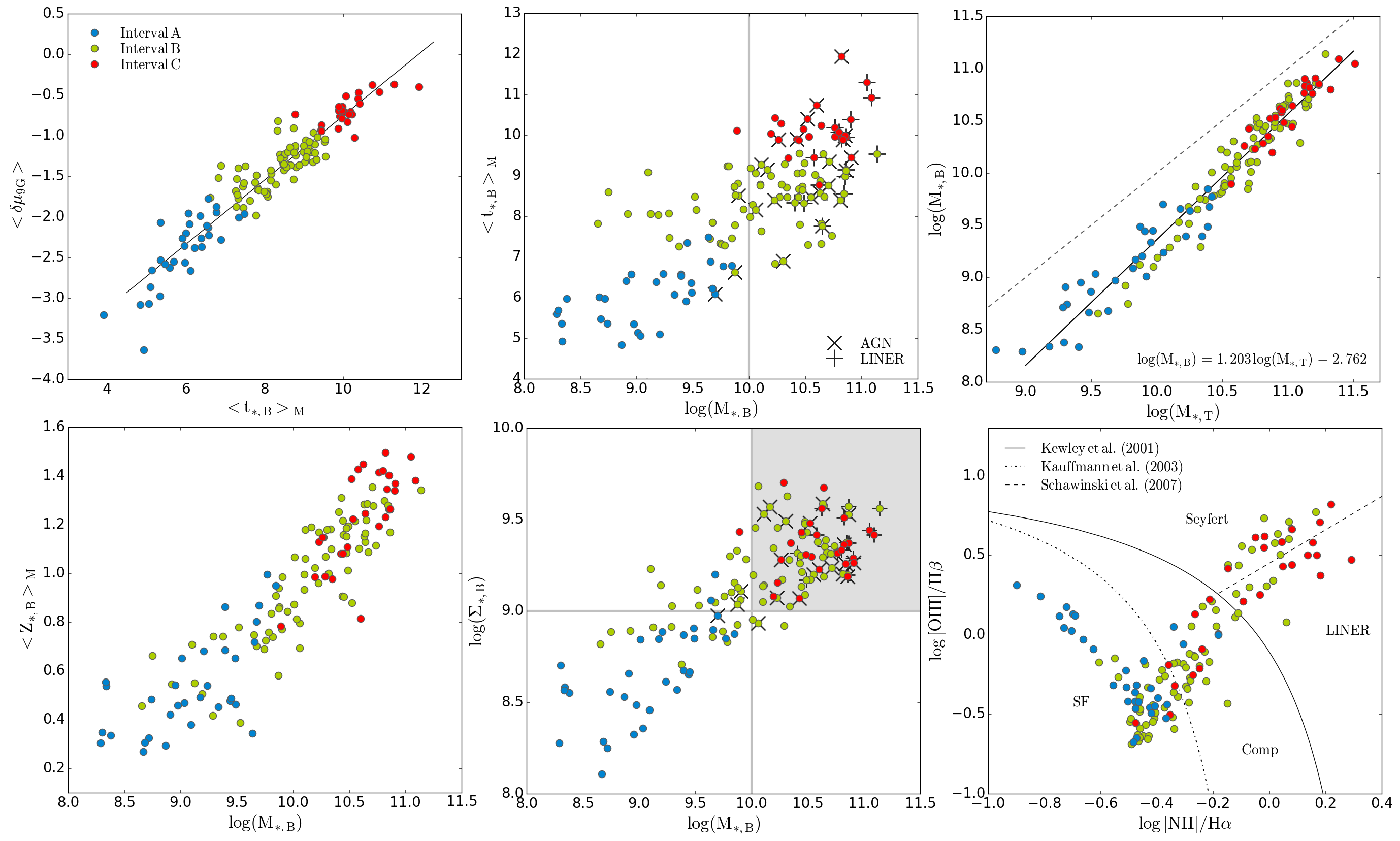}}
\PutLabel{157}{180}{\hvss a}
\PutLabel{200}{296}{\hvss b}
\PutLabel{376}{296}{\hvss c}
\PutLabel{157}{22}{\hvss d}
\PutLabel{330}{22}{\hvss e}
\PutLabel{376}{22}{\hvss f}
\end{picture}
\caption{Main results from the spectral fitting of the LTG sample with the SSP base Z5 (cf. Sect.~\ref{SpectralModeling}). 
The color coding and the meaning of the individual panels is as in Fig,~\ref{res}:
\brem{a)} \dmb\ ($r$ mag) vs. mass-weighted stellar age \mtmass\ (Gyr) for the bulge component, with the solid line showing a linear fit to the data.
\brem{b)} Logarithm of the stellar mass \mbstar\ (\msun) in the bulge vs. mass-weighted stellar age \mtmass\ (Gyr). Galaxies spectroscopically classified as Seyfert and LINER (cf. panel \brem{f}) are marked. 
\brem{c)} Total stellar mass \mstotal\ vs. \mbstar. The solid and dashed lines show, respectively, a linear fit to the data and equality between both quantities. 
\brem{d)} Logarithm of \mbstar\ vs. mass-weighted stellar metallicity \mzmass\ (\zsun).
\brem{e)} Logarithm of \mbstar\ vs. logarithm of the mean stellar surface density \sstar\ (\msun\,kpc$^{-2}$) in the bulge. 
\brem{f)} Spectroscopic classification after BPT line ratios within simulated 3\arcsec\ SDSS fibers (cf. Sect. \ref{BPT}).
}\label{Z5-res}
\end{figure*}
\end{center}      

As it can be seen from comparison of Fig.~\ref{res} with Fig.~\ref{Z5-res}, the AMD is reflected in an increase 
of the metallicity and decrease in the age of high-\mstar\ bulges. However, spectral modeling with the Z5 SSP library 
has a relatively little effect on \dmb\ ($\pm$0.5 mag) and \mstar\ ($\la$0.2-0.3 dex), and, quite importantly, it does not 
appreciably influence the trends between \dmb\ and physical and evolutionary quantities of LTG bulges.\\

\smallskip
For the sake of completeness, we provide in Table~\ref{linfit} a synopsis of relations inferred from regression analysis for quantities obtained from spectral modeling
with the Z4 and Z5 SSP library (panels \brem{a}-\brem{e} in Fig.~\ref{res1} and Fig.~\ref{Z5-res}, respectively). 
The upper block lists relations computed by minimizing $\chi^2$ with regard to both quantities considered, whereas the lower block refers to fits 
where $\chi^2$ minimization was done only with regard to the quantity listed in the right-hand side of each equation.   

\newgeometry{left=3cm}
\pagestyle{empty}

{\tabulinesep=1.1mm

\begin{table}
\begin{tabu}{ p{7cm} p{7cm} }\\

\multicolumn{1}{c}{Z4 SSP Library} & \multicolumn{1}{c}{Z5 SSP Library} \\\hline

\cellcolor[gray]{0.8} \mtmass =  (2.05$\pm$0.04) $\cdot$ \dmb\ + (12.14$\pm$0.06) &
\cellcolor[gray]{0.8} \mtmass =  (2.26$\pm$0.07) $\cdot$ \dmb\ + (11.52$\pm$0.11) \\
\cellcolor[gray]{0.8} \mtmass\ = (2.30$\pm$0.11)$\cdot \log$\,\mbstar\ - (13.44$\pm$1.13) &
\cellcolor[gray]{0.8} \mtmass\ = (1.69$\pm$0.12)$\cdot \log$\,\mbstar\ - (8.71$\pm$1.23) \\
\cellcolor[gray]{0.8} $\log$\,\sstar = (0.42$\pm$0.02)$\cdot \log$\,\mbstar\ + (4.96$\pm$0.23) &
\cellcolor[gray]{0.8} $\log$\,\sstar = (0.39$\pm$0.02)$\cdot \log$\,\mbstar\ + (5.20$\pm$0.24) \\
\cellcolor[gray]{0.8} $\log$\,\sstar = (0.16$\pm$0.01)$\cdot$\mtmass + (7.63$\pm$0.09) & 
\cellcolor[gray]{0.8} $\log$\,\sstar = (0.16$\pm$0.01)$\cdot$\mtmass + (7.76$\pm$0.10) \\
\cellcolor[gray]{0.8} $\log$\,\sstar = (0.32$\pm$0.02)$\cdot$\dmb + (9.56$\pm$0.03) &
\cellcolor[gray]{0.8} $\log$\,\sstar = (0.39$\pm$0.03)$\cdot$\dmb + (9.69$\pm$0.05) \\
\cellcolor[gray]{0.8} \mzmass\  = (0.23$\pm$0.01)$\cdot \log$\,\mbstar\ - (1.60$\pm$0.11) & 
\cellcolor[gray]{0.8} \mzmass\  = (0.42$\pm$0.02)$\cdot \log$\,\mbstar\ - (3.26$\pm$0.17) \\
\cellcolor[gray]{0.8} \mzmass\  = (0.28$\pm$0.01)$\cdot \log$\,\mstotal\ - (2.30$\pm$0.15) & 
\cellcolor[gray]{0.8} \mzmass\  = (0.51$\pm$0.02)$\cdot \log$\,\mstotal\ - (4.48$\pm$0.23) \\
\cellcolor[gray]{0.8} \mdtmass\  = (1.88$\pm$0.12)$\cdot \log$\,\mstotal\ - (11.85$\pm$1.32) & 
\cellcolor[gray]{0.8} \mdtmass\  = (1.42$\pm$0.12)$\cdot \log$\,\mstotal\ - (7.86$\pm$1.25) \\
\cellcolor[gray]{0.8} \mdzmass\  = (0.20$\pm$0.02)$\cdot \log$\,\mstotal\ - (1.55$\pm$0.16) & 
\cellcolor[gray]{0.8} \mdzmass\  = (0.29$\pm$0.02)$\cdot \log$\,\mstotal\ - (2.31$\pm$0.20) \\
\cellcolor[gray]{0.8} $\log$\,\mbstar = (0.67$\pm$0.04)$\cdot$\dmb\ + (10.87$\pm$0.06) &
\cellcolor[gray]{0.8} $\log$\,\mbstar = (0.86$\pm$0.06)$\cdot$\dmb\ + (11.26$\pm$0.09) \\
\cellcolor[gray]{0.8} $\log$\,\mbstar = (1.22$\pm$0.02) $\log$\,\mstotal\ - (2.87$\pm$0.25) &
\cellcolor[gray]{0.8}  $\log$\,\mbstar = (1.20$\pm$0.03) $\log$\,\mstotal\ - (2.67$\pm$0.27) \\
\cellcolor[gray]{0.8} \mbstar/\mstotal\ = (0.13$\pm$0.02)$\cdot$log\,\mstotal\ - (1.12$\pm$0.17) & 
\cellcolor[gray]{0.8} \mbstar/\mstotal\ = (0.13$\pm$0.02)$\cdot$log\,\mstotal\ - (1.09$\pm$0.18) \\
\cellcolor[gray]{0.8} \mtmass\  = (1.21$\pm$0.06)$\cdot$\mdtmass\  - (0.18$\pm$0.49) & 
\cellcolor[gray]{0.8} \mtmass\  = (1.07$\pm$0.08)$\cdot$\mdtmass\  + (0.64$\pm$0.55) \\
\cellcolor[gray]{0.8} \mzmass\  = (1.09$\pm$0.05)$\cdot$\mdzmass\  + (0.05$\pm$0.03) &  
\cellcolor[gray]{0.8} \mzmass\  = (1.42$\pm$0.06)$\cdot$\mdzmass\  - (0.17$\pm$0.04) \\
\cellcolor[gray]{0.8} \dbdt\  = (1.00$\pm$0.12)$\cdot$log\,\mstotal\ - (9.07$\pm$1.28) & 
\cellcolor[gray]{0.8} \dbdt\  = (0.74$\pm$0.14)$\cdot$log\,\mstotal\ - (6.65$\pm$1.45) \\
\cellcolor[gray]{0.8} \dbdz\  = (0.08$\pm$0.01)$\cdot$log\,\mstotal\ - (0.75$\pm$0.12) &  
\cellcolor[gray]{0.8} \dbdz\  = (0.22$\pm$0.02)$\cdot$log\,\mstotal\ - (2.18$\pm$0.16) \\
\hline
\end{tabu}
\begin{tabu}{ p{7cm} p{7cm} }\\
\hline
\cellcolor[gray]{0.8} \mtmass =  (2.13$\pm$0.04) $\cdot$ \dmb\ + (12.23$\pm$0.23) & 
\cellcolor[gray]{0.8} \mtmass =  (2.49$\pm$0.07) $\cdot$ \dmb\ + (11.86$\pm$0.37)  \\ 
\cellcolor[gray]{0.8} \mtmass\ = (2.93$\pm$0.15)$\cdot \log$\,\mbstar\ - (19.81$\pm$1.06) & 
\cellcolor[gray]{0.8} \mtmass\ = (2.66$\pm$0.20)$\cdot \log$\,\mbstar\ - (18.39$\pm$1.45) \\  
\cellcolor[gray]{0.8} $\log$\,\sstar = (0.44$\pm$0.03)$\cdot \log$\,\mbstar\ + (4.70$\pm$0.30) &
\cellcolor[gray]{0.8} $\log$\,\sstar = (0.42$\pm$0.03)$\cdot \log$\,\mbstar\ + (4.93$\pm$0.31)   \\
\cellcolor[gray]{0.8} $\log$\,\sstar = (0.16$\pm$0.01)$\cdot$\mtmass + (7.62$\pm$0.10) & 
\cellcolor[gray]{0.8} $\log$\,\sstar = (0.17$\pm$0.01)$\cdot$\mtmass + (7.74$\pm$0.12) \\
\cellcolor[gray]{0.8} $\log$\,\sstar = (0.34$\pm$0.02)$\cdot$\dmb + (9.58$\pm$0.07) &
\cellcolor[gray]{0.8} $\log$\,\sstar = (0.43$\pm$0.03)$\cdot$\dmb + (9.75$\pm$0.14) \\
\cellcolor[gray]{0.8} \mzmass\  = (0.24$\pm$0.01)$\cdot \log$\,\mbstar\ - (1.64$\pm$0.11) &
\cellcolor[gray]{0.8} \mzmass\  = (0.43$\pm$0.02)$\cdot \log$\,\mbstar\ - (3.40$\pm$0.15) \\
\cellcolor[gray]{0.8} \mzmass\  = (0.28$\pm$0.02)$\cdot \log$\,\mstotal\ - (2.37$\pm$0.15) &
\cellcolor[gray]{0.8} \mzmass\  = (0.54$\pm$0.02)$\cdot \log$\,\mstotal\ - (4.76$\pm$0.20) \\
\cellcolor[gray]{0.8} \mdtmass\  = (2.80$\pm$0.19)$\cdot \log$\,\mstotal\ - (21.56$\pm$1.48) & 
\cellcolor[gray]{0.8} \mdtmass\  = (2.44$\pm$0.21)$\cdot \log$\,\mstotal\ - (18.64$\pm$1.65) \\
\cellcolor[gray]{0.8} \mdzmass\  = (0.21$\pm$0.02)$\cdot \log$\,\mstotal\ - (1.61$\pm$0.16) &
\cellcolor[gray]{0.8} \mdzmass\  = (0.31$\pm$0.02)$\cdot \log$\,\mstotal\ - (2.45$\pm$0.20) \\
\cellcolor[gray]{0.8} $\log$\,\mbstar = (0.76$\pm$0.04)$\cdot$\dmb\ + (10.98$\pm$0.30) &
\cellcolor[gray]{0.8} $\log$\,\mbstar = (1.11$\pm$0.08)$\cdot$\dmb\ + (11.63$\pm$0.57) \\
\cellcolor[gray]{0.8} $\log$\,\mbstar = (1.26$\pm$0.02) $\log$\,\mstotal\ - (3.27$\pm$0.23) & 
\cellcolor[gray]{0.8} $\log$\,\mbstar = (1.25$\pm$0.03) $\log$\,\mstotal\ - (3.12$\pm$0.24) \\
\cellcolor[gray]{0.8} \mbstar/\mstotal\ = (0.14$\pm$0.02)$\cdot$log\,\mstotal\ - (1.17$\pm$0.17) & 
\cellcolor[gray]{0.8} \mbstar/\mstotal\ = (0.14$\pm$0.02)$\cdot$log\,\mstotal\ - (1.15$\pm$0.19) \\
\cellcolor[gray]{0.8} \mtmass\  = (1.46$\pm$0.07)$\cdot$\mdtmass\  - (2.23$\pm$0.54) &
\cellcolor[gray]{0.8} \mtmass\  = (1.52$\pm$0.12)$\cdot$\mdtmass\  - (2.56$\pm$0.70) \\
\cellcolor[gray]{0.8} \mzmass\  = (1.26$\pm$0.06)$\cdot$\mdzmass\  - (0.04$\pm$0.03) & 
\cellcolor[gray]{0.8} \mzmass\  = (1.63$\pm$0.07)$\cdot$\mdzmass\  - (0.33$\pm$0.04) \\
\cellcolor[gray]{0.8} \dbdt\  = (2.35$\pm$0.31)$\cdot$log\,\mstotal\ - (23.42$\pm$2.68) & 
\cellcolor[gray]{0.8} \dbdt\  = (3.11$\pm$0.64)$\cdot$log\,\mstotal\ - (31.57$\pm$6.00) \\
\cellcolor[gray]{0.8} \dbdz\  = (0.08$\pm$0.01)$\cdot$log\,\mstotal\ - (0.76$\pm$0.12) & 
\cellcolor[gray]{0.8} \dbdz\  = (0.23$\pm$0.01)$\cdot$log\,\mstotal\ - (2.25$\pm$0.16) \\\hline

\end{tabu}
\vspace{0.1cm}
\caption{Overview of the fits obtained for quantities inferred from analysis of the LTG sample with the Z4 and Z5 SSP library. 
The \dmb\ is expressed in SDSS $r$ mag, \mtmass\ and \mdtmass\ in Gyr, \mbstar\ and \mstotal\ in \msun, \sstar\ in $10^9$ \msun\,kpc$^{-2}$ and \mzmass\ in \zsun. \label{linfit}}
\end{table}



\newgeometry{left=2.0cm,right=2.0cm,top=0.1cm,bottom=0.1cm}
\begin{landscape}

\begin{changemargin}{2.3cm}{0.5cm}
\section{Photometric, evolutionary and physical quantities for the galaxy sample \label{A-table}}
\end{changemargin}

\begin{longtable}{ccccccccccccccc}

\setlength\LTcapleft{11.5cm}\\
\caption[\textbf{Table C.1}]{Photometric, evolutionary and physical quantities for the galaxy sample under study. \brem{Col. 1:} name of the galaxy, \brem{Col. 2:} assumed distance in Mpc, \brem{Col. 3:} isophotal radius of the bulge (\rbulge) in kpc, \brem{Col. 4:} logarithm of the total stellar mass (\mstotal\ in \msun), \brem{Col. 5:} logarithm of the present-day stellar mass (\mbstar\ in \msun) within \rbulge, \brem{Cols. 6\&7:} mass-weighted stellar age of the bulge (\mtmass\ in Gyr) and its dispersion ($\sigma$ \mtmass), \brem{Cols. 8\&9:} mass-weighted stellar age of the disk (\mdtmass\ in Gyr) and its dispersion ($\sigma$ \mdtmass), \brem{Cols. 10\&11:} mass-weighted stellar metallicity of the bulge (\mzmass\ in \zsun) and its dispersion ($\sigma$ \mzmass), \brem{Cols. 12\&13:} mass-weighted stellar metallicity of the disk (\mdzmass\ in \zsun) and its dispersion ($\sigma$ \mzmass), \brem{Col. 14:} logarithm of the mean stellar mass surface density in the bulge (\sstar\ in \msun\,kpc$^{-2}$) and \brem{Col. 15:} mean \dmb\ (\textit{r} mag) within \rbulge\. Quantities listed in cols. 4--15 were obtained from spectral modeling with \Starlight\ of CALIFA V500 data with the Z4 SSP library and refer to the present-day \mstar\ (cf. Sect. \ref{SpectralModeling}).}\label{tab1}\\

Name & \multicolumn{1}{p{1cm}}{\centering Distance \\ (Mpc) \\ (2)} &
\multicolumn{1}{p{1cm}}{\centering \rbulge\ \\ (kpc) \\ (3)}   &
\multicolumn{1}{p{1.5cm}}{\centering log \mstotal\ \\ (\msun) \\(4)} &
\multicolumn{1}{p{1.5cm}}{\centering log \mbstar\ \\ (\msun) \\ (5)} &
\multicolumn{1}{p{1cm}}{\centering \mtmass\ \\ (Gyr) \\ (6)} &
\multicolumn{1}{p{1.5cm}}{\centering $\sigma$ \mtmass\ \\ (Gyr) \\ (7)} &
\multicolumn{1}{p{1cm}}{\centering \mdtmass\ \\ (Gyr) \\ (8)} &
\multicolumn{1}{p{1.5cm}}{\centering $\sigma$ \mdtmass\ \\ (Gyr) \\ (9)} &
\multicolumn{1}{p{1cm}}{\centering \mzmass\ \\ (\zsun) \\ (10)} &
\multicolumn{1}{p{1.5cm}}{\centering $\sigma$ \mzmass\ \\ (\zsun) \\ (11)} &
\multicolumn{1}{p{1cm}}{\centering \mdzmass\ \\ (\zsun) \\ (12)} &
\multicolumn{1}{p{1.5cm}}{\centering $\sigma$ \mdzmass\ \\ (\zsun) \\ (13)} &
\multicolumn{1}{p{1.5cm}}{\centering $\log\,\,$ \sstar\ \\ (\msun\,kpc$^{-2}$) \\ (14)} & \multicolumn{1}{p{1cm}}{\centering \dmb\ \\
(mag) \\ (15)}  \\

\rule[0.0ex]{0cm}{2.5ex}
CGCG163\_062 & 68.1 & 1.37 & 9.95 & 9.31 & 6.43 & 0.89 & 4.80 & 0.76 & 0.37 &
0.22 & 0.34 & 0.22 & 8.54 & -2.64 \\

IC0159 & 52.0 & 0.87 & 10.12 & 9.27 & 8.28 & 1.14 & 6.67 & 1.19 & 0.35 &
0.16 & 0.38 & 0.19 & 8.90 & -1.97 \\

IC0208 & 46.6 & 1.22 & 10.25 & 9.44 & 9.11 & 0.99 & 7.68 & 1.01 & 0.53 &
0.24 & 0.62 & 0.31 & 8.77 & -1.20 \\

IC0776 & 40.2 & 1.07 & 9.58 & 8.71 & 6.49 & 1.02 & 7.41 & 1.12 & 0.25 & 0.10
& 0.30 & 0.16 & 8.15 & -2.67 \\

IC1256 & 72.1 & 1.44 & 10.93 & 10.13 & 10.85 & 0.67 & 8.67 & 1.03 & 0.77 &
0.32 & 0.68 & 0.31 & 9.32 & -0.63 \\

IC1528 & 51.6 & 1.01 & 10.52 & 9.83 & 10.51 & 0.71 & 6.95 & 0.98 & 0.47 &
0.18 & 0.44 & 0.22 & 9.33 & -0.94 \\

IC1683 & 65.4 & 1.91 & 10.85 & 10.53 & 11.16 & 0.73 & 8.28 & 0.99 & 0.71 &
0.32 & 0.72 & 0.36 & 9.47 & -0.60 \\

IC2604 & 28.9 & 0.46 & 9.26 & 8.37 & 6.42 & 0.97 & 6.81 & 0.99 & 0.33 &
0.15 & 0.38 & 0.27 & 8.54 & -2.75 \\

IC3918 & 97.7 & 1.17 & 9.85 & 9.09 & 4.99 & 1.27 & 6.13 & 1.11 & 0.30 & 0.14
& 0.36 & 0.22 & 8.45 & -3.48 \\

IC4566 & 86.3 & 2.97 & 11.16 & 10.79 & 12.03 & 0.50 & 9.81 & 0.82 & 0.78 &
0.47 & 0.72 & 0.38 & 9.35 & -0.20 \\

MCG01\_10\_019 & 69.9 & 1.63 & 10.43 & 9.86 & 9.24 & 0.89 & 8.78 & 0.92 &
0.75 & 0.33 & 0.67 & 0.33 & 8.94 & -1.27 \\

MCG09\_22\_053 & 55.0 & 1.18 & 9.89 & 9.50 & 6.55 & 0.94 & 6.56 & 1.00 & 0.41 &
0.15 & 0.34 & 0.16 & 8.86 & -2.98 \\

NGC0001 & 61.6 & 1.92 & 10.99 & 10.68 & 9.76 & 0.97 & 8.12 & 0.98 & 0.77 &
0.30 & 0.59 & 0.27 & 9.62 & -1.18 \\

NGC0036 & 81.0 & 2.52 & 11.22 & 10.76 & 10.99 & 0.93 & 9.11 & 0.91 & 0.86 &
0.50 & 0.72 & 0.36 & 9.46 & -0.61 \\

NGC0160 & 70.5 & 3.14 & 11.18 & 10.85 & 11.54 & 0.63 & 8.86 & 0.80 & 0.88 &
0.57 & 0.85 & 0.57 & 9.36 & -0.44 \\

NGC0165 & 78.9 & 2.00 & 10.85 & 10.29 & 9.97 & 0.76 & 9.01 & 0.98 & 0.68 &
0.34 & 0.58 & 0.27 & 9.19 & -0.86 \\

NGC0171 & 52.8 & 1.98 & 10.92 & 10.47 & 11.44 & 0.76 & 9.00 & 0.90 & 0.76 &
0.51 & 0.65 & 0.34 & 9.38 & -0.40 \\

NGC0180 & 70.6 & 2.17 & 11.14 & 10.84 & 11.51 & 0.63 & 10.39 & 0.84 & 0.83
& 0.49 & 0.82 & 0.39 & 9.67 & -0.44 \\

NGC0214 & 61.0 & 1.88 & 11.16 & 10.68 & 11.43 & 0.65 & 9.21 & 0.85 & 0.79 &
0.40 & 0.64 & 0.39 & 9.64 & -0.51 \\

NGC0234 & 59.7 & 2.33 & 11.00 & 10.48 & 10.54 & 0.89 & 8.67 & 1.04 & 0.73 &
0.32 & 0.62 & 0.27 & 9.25 & -0.92 \\

NGC0237 & 55.9 & 1.22 & 10.58 & 9.97 & 9.81 & 0.99 & 7.92 & 1.01 & 0.53 &
0.23 & 0.41 & 0.19 & 9.30 & -1.03 \\

NGC0257 & 70.4 & 2.66 & 11.13 & 10.69 & 10.24 & 0.87 & 8.36 & 0.88 & 0.78 &
0.33 & 0.60 & 0.34 & 9.34 & -0.98 \\

NGC0309 & 75.8 & 2.78 & 11.1 & 10.55 & 10.99 & 0.72 & 9.29 & 0.95 & 0.77 &
0.48 & 0.68 & 0.30 & 9.16 & -0.61 \\

NGC0447 & 75.1 & 3.81 & 11.53 & 11.24 & 11.21 & 0.62 & 9.14 & 1.00 & 0.93 &
0.60 & 0.80 & 0.43 & 9.59 & -0.57 \\

NGC0477 & 79.1 & 1.89 & 10.78 & 10.29 & 9.99 & 0.83 & 8.52 & 0.95 & 0.71 &
0.33 & 0.57 & 0.23 & 9.24 & -0.99 \\

NGC0496 & 80.5 & 1.48 & 10.72 & 9.88 & 9.86 & 0.73 & 8.42 & 0.97 & 0.51 &
0.19 & 0.42 & 0.19 & 9.04 & -1.13 \\

NGC0768 & 93.3 & 1.90 & 10.89 & 10.35 & 11.07 & 0.64 & 8.14 & 0.98 & 0.74 &
0.34 & 0.48 & 0.21 & 9.30 & -0.55 \\

NGC0776 & 65.5 & 2.44 & 11.09 & 10.76 & 11.57 & 0.58 & 9.29 & 0.94 & 0.86 &
0.44 & 0.67 & 0.31 & 9.48 & -0.30 \\

NGC0873 & 53.2 & 1.10 & 10.76 & 10.22 & 10.95 & 0.76 & 8.35 & 1.11 & 0.51 &
0.23 & 0.47 & 0.23 & 9.63 & -0.92 \\

NGC0941 & 21.3 & 0.47 & 9.48 & 8.64 & 8.02 & 0.97 & 6.48 & 1.08 & 0.41 &
0.15 & 0.36 & 0.19 & 8.81 & -1.97 \\

NGC0976 & 57.0 & 2.49 & 11.05 & 10.86 & 10.23 & 0.73 & 7.91 & 0.78 & 0.78 &
0.44 & 0.62 & 0.42 & 9.58 & -0.94 \\

NGC0991 & 20.2 & 0.70 & 9.49 & 8.81 & 5.69 & 0.94 & 6.23 & 1.07 & 0.43 &
0.19 & 0.39 & 0.19 & 8.62 & -3.16 \\

NGC1070 & 54.0 & 2.55 & 11.23 & 10.91 & 12.07 & 0.48 & 10.82 & 0.70 & 0.83 &
0.48 & 0.76 & 0.39 & 9.59 & -0.18 \\

NGC1093 & 70.8 & 1.10 & 10.85 & 10.19 & 11.58 & 0.57 & 9.46 & 0.93 & 0.89 &
0.49 & 0.69 & 0.29 & 9.61 & -0.29 \\

NGC1094 & 85.8 & 2.77 & 11.06 & 10.74 & 9.18 & 0.84 & 9.01 & 0.87 & 0.77 &
0.37 & 0.53 & 0.30 & 9.36 & -1.43 \\

NGC1590 & 52.2 & 0.87 & 10.6 & 10.03 & 10.29 & 0.94 & 7.57 & 0.89 & 0.51 &
0.21 & 0.39 & 0.22 & 9.65 & -1.02 \\

NGC1645 & 65.9 & 1.92 & 11.04 & 10.71 & 11.18 & 0.93 & 9.52 & 0.73 & 0.93 &
0.71 & 0.71 & 0.41 & 9.65 & -0.54 \\

NGC1659 & 61.7 & 1.38 & 10.73 & 10.2 & 9.38 & 0.95 & 8.03 & 1.04 & 0.66 &
0.38 & 0.55 & 0.34 & 9.42 & -1.39 \\

NGC1667 & 61.2 & 2.63 & 11.27 & 10.91 & 11.4 & 0.70 & 9.61 & 0.84 & 0.74 &
0.37 & 0.54 & 0.27 & 9.57 & -0.55 \\

NGC2253 & 51.3 & 1.25 & 10.9 & 10.37 & 10.84 & 0.79 & 9.35 & 0.95 & 0.66 &
0.25 & 0.52 & 0.20 & 9.68 & -0.58 \\

NGC2347 & 63.0 & 1.72 & 11.12 & 10.73 & 11.59 & 0.64 & 10.13 & 0.88 & 0.85
& 0.43 & 0.59 & 0.25 & 9.77 & -0.36 \\

NGC2486 & 66.3 & 1.87 & 10.93 & 10.5 & 12.00 & 0.45 & 10.28 & 0.67 & 0.89 &
0.69 & 0.77 & 0.45 & 9.45 & -0.23 \\

NGC2487 & 68.9 & 2.68 & 11.07 & 10.66 & 11.07 & 0.76 & 9.47 & 0.90 & 0.84 &
0.45 & 0.67 & 0.32 & 9.3 & -0.62 \\

NGC2530 & 71.7 & 1.51 & 10.55 & 9.90 & 9.01 & 1.01 & 6.91 & 1.05 & 0.5 &
0.22 & 0.36 & 0.17 & 9.04 & -1.58 \\

NGC2540 & 89.0 & 1.72 & 10.79 & 10.13 & 9.69 & 0.85 & 7.77 & 1.00 & 0.64 &
0.31 & 0.48 & 0.24 & 9.16 & -1.35 \\

NGC2543 & 37.4 & 1.5 & 10.58 & 10.23 & 10.55 & 0.71 & 8.35 & 0.97 & 0.67 &
0.36 & 0.53 & 0.25 & 9.38 & -0.88 \\

NGC2558 & 71.6 & 2.15 & 11.01 & 10.65 & 11.69 & 0.62 & 9.51 & 0.79 & 0.91 &
0.62 & 0.81 & 0.49 & 9.48 & -0.34 \\

NGC2595 & 62.7 & 2.35 & 11.00 & 10.58 & 11.6 & 0.56 & 9.50 & 0.87 & 0.78 &
0.41 & 0.66 & 0.31 & 9.34 & -0.36 \\

NGC2604 & 32.3 & 0.59 & 9.91 & 8.97 & 8.86 & 0.86 & 6.49 & 0.98 & 0.31 &
0.17 & 0.34 & 0.21 & 8.94 & -1.87 \\

NGC2730 & 56.7 & 1.04 & 10.44 & 9.47 & 7.42 & 0.98 & 6.87 & 0.98 & 0.63 &
0.19 & 0.43 & 0.20 & 8.94 & -2.14 \\

NGC2780 & 31.8 & 0.49 & 9.98 & 9.09 & 9.67 & 0.88 & 7.84 & 0.98 & 0.52 &
0.18 & 0.53 & 0.26 & 9.22 & -0.98 \\

NGC2805 & 28.7 & 0.69 & 9.99 & 9.13 & 5.87 & 0.93 & 6.96 & 0.97 & 0.57 &
0.21 & 0.46 & 0.25 & 8.96 & -2.90 \\

NGC2906 & 33.5 & 1.11 & 10.68 & 10.23 & 11.80 & 0.51 & 9.66 & 0.92 & 0.85 &
0.43 & 0.60 & 0.27 & 9.64 & -0.20 \\

NGC2916 & 56.0 & 2.86 & 11.00 & 10.61 & 9.09 & 0.98 & 8.84 & 0.97 & 0.81 &
0.45 & 0.52 & 0.25 & 9.20 & -1.14 \\

NGC3057 & 25.9 & 0.96 & 9.43 & 8.72 & 5.96 & 0.93 & 6.41 & 1.07 & 0.28 &
0.11 & 0.32 & 0.16 & 8.26 & -3.03 \\

NGC3381 & 28.8 & 0.72 & 9.91 & 9.11 & 8.42 & 0.98 & 6.95 & 1.14 & 0.41 &
0.17 & 0.40 & 0.17 & 8.90 & -2.02 \\

NGC3614 & 38.4 & 1.41 & 10.37 & 9.74 & 8.64 & 1.16 & 7.87 & 1.07 & 0.59 &
0.23 & 0.51 & 0.22 & 8.95 & -1.39 \\

NGC3687 & 41.1 & 2.13 & 10.55 & 10.27 & 9.96 & 1.10 & 8.74 & 0.97 & 0.73 &
0.34 & 0.59 & 0.25 & 9.11 & -0.84 \\

NGC3811 & 49.0 & 2.08 & 10.74 & 10.43 & 9.88 & 1.06 & 7.70 & 1.07 & 0.60 &
0.25 & 0.49 & 0.22 & 9.29 & -0.97 \\

NGC3913 & 19.2 & 0.75 & 9.34 & 8.91 & 6.91 & 0.98 & 6.81 & 1.03 & 0.38 &
0.16 & 0.39 & 0.20 & 8.67 & -2.83 \\

NGC4047 & 53.3 & 2.28 & 10.99 & 10.63 & 10.22 & 0.88 & 9.39 & 0.96 & 0.73 &
0.29 & 0.55 & 0.21 & 9.42 & -0.91 \\

NGC4185 & 61.0 & 2.04 & 10.95 & 10.26 & 11.22 & 0.64 & 10.43 & 0.79 & 0.79
& 0.40 & 0.80 & 0.40 & 9.15 & -0.38 \\

NGC4210 & 43.2 & 1.41 & 10.57 & 9.94 & 9.80 & 0.95 & 8.8 & 1.02 & 0.75 &
0.33 & 0.61 & 0.23 & 9.15 & -0.88 \\

NGC4711 & 63.3 & 1.25 & 10.82 & 9.86 & 9.77 & 0.98 & 7.93 & 0.94 & 0.64 &
0.25 & 0.46 & 0.22 & 9.17 & -1.02 \\

NGC4961 & 42.5 & 0.60 & 10.01 & 9.19 & 8.87 & 1.02 & 7.13 & 1.06 & 0.39 &
0.14 & 0.29 & 0.14 & 9.14 & -1.54 \\

NGC5000 & 84.9 & 2.18 & 11.02 & 10.54 & 11.57 & 0.52 & 9.87 & 0.84 & 0.75 &
0.44 & 0.64 & 0.32 & 9.36 & -0.25 \\

NGC5016 & 43.5 & 0.96 & 10.59 & 9.91 & 10.4 & 0.76 & 8.73 & 0.89 & 0.64 &
0.28 & 0.53 & 0.28 & 9.45 & -0.77 \\

NGC5056 & 84.6 & 1.74 & 10.91 & 10.4 & 11.14 & 0.66 & 8.18 & 0.98 & 0.63 &
0.29 & 0.48 & 0.20 & 9.42 & -0.50 \\

NGC5157 & 108.0 & 3.84 & 11.32 & 10.93 & 12.48 & 0.47 & 9.72 & 0.86 & 0.89
& 0.69 & 0.78 & 0.40 & 9.26 & -0.1 \\

NGC5205 & 30.9 & 1.30 & 10.15 & 9.79 & 8.49 & 1.12 & 7.44 & 0.94 & 0.64 &
0.30 & 0.44 & 0.23 & 9.06 & -1.57 \\

NGC5320 & 43.6 & 1.34 & 10.59 & 9.96 & 10.57 & 0.84 & 8.85 & 1.00 & 0.66 &
0.27 & 0.51 & 0.21 & 9.21 & -0.66 \\

NGC5378 & 49.6 & 2.69 & 10.79 & 10.49 & 11.40 & 0.71 & 9.98 & 0.76 & 0.78 &
0.46 & 0.73 & 0.42 & 9.13 & -0.34 \\

NGC5406 & 79.0 & 3.61 & 11.34 & 11.02 & 12.09 & 0.50 & 9.94 & 0.80 & 0.93 &
0.61 & 0.73 & 0.35 & 9.4 & -0.24 \\

NGC5480 & 32.9 & 1.23 & 10.35 & 9.76 & 7.97 & 0.94 & 7.94 & 1.06 & 0.50 &
0.23 & 0.44 & 0.22 & 9.09 & -1.95 \\

NGC5519 & 111.3 & 2.64 & 11.20 & 10.93 & 10.76 & 0.56 & 8.32 & 0.98 & 0.70 &
0.33 & 0.65 & 0.28 & 9.59 & -0.83 \\

NGC5622 & 60.2 & 1.29 & 10.56 & 9.91 & 11.09 & 0.69 & 8.96 & 0.99 & 0.6 &
0.27 & 0.52 & 0.21 & 9.19 & -0.43 \\

NGC5633 & 39.6 & 1.10 & 10.63 & 10.12 & 10.23 & 0.86 & 8.61 & 1.00 & 0.58 &
0.22 & 0.39 & 0.18 & 9.54 & -1.00 \\

NGC5656 & 51.4 & 1.86 & 10.86 & 10.49 & 9.99 & 0.95 & 8.40 & 0.95 & 0.64 &
0.23 & 0.41 & 0.17 & 9.45 & -0.85 \\

NGC5665 & 37.3 & 1.05 & 10.52 & 9.81 & 9.33 & 0.91 & 7.89 & 1.14 & 0.37 &
0.15 & 0.43 & 0.19 & 9.27 & -1.51 \\

NGC5720 & 113.0 & 3.09 & 11.20 & 10.75 & 11.36 & 1.05 & 8.78 & 0.97 & 0.85 &
0.55 & 0.68 & 0.31 & 9.27 & -0.49 \\

NGC5732 & 59.3 & 1.17 & 10.29 & 9.71 & 9.54 & 0.88 & 7.36 & 0.94 & 0.53 &
0.18 & 0.33 & 0.15 & 9.08 & -1.23 \\

NGC5735 & 59.6 & 2.06 & 10.68 & 10.19 & 10.24 & 0.90 & 7.60 & 1.04 & 0.54 &
0.25 & 0.50 & 0.24 & 9.06 & -0.84 \\

NGC5772 & 74.8 & 3.62 & 11.2 & 10.94 & 12.2 & 0.46 & 10.13 & 0.79 & 0.87 &
0.59 & 0.70 & 0.35 & 9.33 & -0.19 \\

NGC5829 & 85.8 & 1.68 & 10.65 & 9.99 & 9.6 & 1.00 & 7.68 & 1.04 & 0.72 &
0.33 & 0.52 & 0.27 & 9.04 & -1.3 \\

NGC5888 & 126.7 & 3.59 & 11.48 & 10.88 & 11.19 & 0.92 & 8.49 & 1.00 & 0.99 & 0.32 & 0.88 & 0.31 & 9.27 & -0.44 \\

NGC5947 & 88.0 & 2.55 & 10.85 & 10.5 & 10.83 & 0.99 & 8.04 & 1.05 & 0.80 &
0.38 & 0.63 & 0.24 & 9.19 & -0.62 \\

NGC5950 & 43.3 & 1.00 & 10.05 & 9.54 & 8.77 & 0.84 & 7.85 & 1.06 & 0.34 &
0.13 & 0.42 & 0.29 & 9.04 & -1.72 \\

NGC5957 & 32.0 & 1.48 & 10.43 & 9.95 & 8.80 & 1.19 & 8.78 & 0.92 & 0.72 &
0.37 & 0.75 & 0.38 & 9.11 & -1.24 \\

NGC6004 & 60.8 & 2.56 & 10.95 & 10.45 & 10.82 & 0.71 & 8.61 & 0.93 & 0.72 &
0.34 & 0.59 & 0.26 & 9.13 & -0.69 \\

NGC6032 & 67.0 & 1.71 & 10.73 & 10.17 & 10.72 & 0.87 & 8.90 & 0.90 & 0.63 &
0.35 & 0.67 & 0.42 & 9.21 & -0.59 \\

NGC6063 & 46.7 & 0.76 & 10.36 & 9.34 & 9.01 & 1.02 & 7.58 & 1.03 & 0.60 &
0.20 & 0.50 & 0.22 & 9.08 & -1.05 \\

NGC6154 & 88.7 & 3.47 & 11.21 & 10.86 & 11.86 & 0.60 & 9.61 & 0.93 & 0.92 &
0.68 & 0.80 & 0.40 & 9.29 & -0.28 \\

NGC6155 & 40.3 & 0.98 & 10.44 & 9.74 & 7.91 & 1.04 & 6.29 & 1.01 & 0.61 &
0.19 & 0.44 & 0.20 & 9.26 & -1.74 \\

NGC6301 & 120.3 & 2.74 & 11.32 & 10.62 & 10.36 & 0.66 & 8.63 & 0.89 & 0.74
& 0.42 & 0.63 & 0.34 & 9.25 & -0.88 \\

NGC6941 & 88.6 & 2.97 & 11.31 & 10.81 & 12.08 & 0.57 & 9.89 & 0.79 & 0.87 &
0.62 & 0.73 & 0.43 & 9.37 & -0.2 \\

NGC7321 & 97.9 & 3.07 & 11.32 & 11.02 & 11.36 & 0.68 & 8.62 & 0.96 & 0.87 &
0.47 & 0.61 & 0.27 & 9.55 & -0.42 \\

NGC7489 & 85.1 & 1.75 & 11.18 & 10.48 & 10.07 & 0.80 & 8.03 & 0.88 & 0.55 &
0.25 & 0.44 & 0.23 & 9.49 & -1.19 \\

NGC7653 & 58.3 & 2.68 & 10.84 & 10.55 & 8.94 & 0.91 & 7.27 & 0.98 & 0.61 &
0.24 & 0.47 & 0.21 & 9.20 & -1.39 \\

NGC7691 & 55.2 & 1.10 & 10.52 & 9.50 & 7.10 & 0.87 & 7.32 & 0.71 & 0.63 & 0.32
& 0.67 & 0.49 & 8.92 & -2.56 \\

NGC7716 & 35.6 & 1.44 & 10.64 & 10.34 & 10.40 & 1.08 & 9.17 & 0.92 & 0.76 &
0.36 & 0.50 & 0.22 & 9.53 & -0.73 \\

NGC7738 & 91.4 & 3.47 & 11.34 & 11.12 & 11.66 & 0.60 & 9.46 & 0.81 & 0.74 &
0.45 & 0.78 & 0.43 & 9.55 & -0.40 \\

NGC7782 & 72.6 & 3.87 & 11.42 & 11.11 & 12.09 & 0.32 & 10.26 & 0.70 & 0.90 &
0.74 & 0.66 & 0.38 & 9.44 & -0.21 \\

NGC7800 & 24.7 & 0.83 & 9.60 & 8.88 & 5.28 & 1.15 & 4.79 & 1.05 & 0.22 & 0.10
& 0.24 & 0.13 & 8.54 & -3.78 \\

NGC7819 & 67.2 & 1.91 & 10.46 & 9.99 & 9.03 & 0.93 & 6.94 & 1.02 & 0.50 &
0.22 & 0.46 & 0.20 & 8.93 & -1.5 \\

UGC00312 & 58.5 & 1.10 & 10.31 & 9.60 & 7.53 & 0.96 & 6.17 & 1.04 & 0.32 &
0.14 & 0.28 & 0.15 & 9.02 & -2.98 \\

UGC02311 & 94.5 & 3.01 & 10.97 & 10.60 & 9.19 & 0.85 & 7.80 & 0.95 & 0.72 &
0.42 & 0.56 & 0.29 & 9.15 & -1.11 \\

UGC02443 & 33.2 & 0.48 & 9.85 & 8.97 & 9.53 & 1.08 & 7.55 & 0.94 & 0.58 &
0.24 & 0.42 & 0.21 & 9.10 & -1.02 \\

UGC03253 & 59.5 & 1.77 & 10.96 & 10.71 & 10.11 & 0.79 & 9.12 & 0.91 & 0.79
& 0.49 & 0.61 & 0.27 & 9.71 & -0.69 \\

UGC03973 & 93.1 & 3.02 & 11.11 & 10.68 & 11.59 & 0.51 & 10.33 & 0.71 & 0.50
& 0.33 & 0.50 & 0.27 & 9.23 & -0.1 \\

UGC04195 & 69.8 & 1.94 & 10.82 & 10.28 & 12.02 & 0.44 & 9.81 & 0.87 & 0.75
& 0.46 & 0.72 & 0.36 & 9.21 & -0.15 \\

UGC04262 & 80.8 & 2.75 & 10.96 & 10.58 & 10.79 & 0.84 & 10.29 & 0.81 & 0.80
& 0.44 & 0.70 & 0.35 & 9.20 & -0.48 \\

UGC04308 & 52.1 & 1.73 & 11.07 & 10.3 & 9.04 & 0.90 & 8.05 & 1.03 & 0.64 &
0.32 & 0.54 & 0.25 & 9.32 & -1.59 \\

UGC04375 & 31.7 & 0.90 & 10.41 & 9.76 & 8.81 & 0.82 & 8.11 & 0.90 & 0.73 &
0.37 & 0.63 & 0.30 & 9.36 & -1.57 \\

UGC04455 & 130.0 & 3.05 & 11.3 & 10.78 & 11.85 & 0.43 & 8.21 & 0.64 & 0.86
& 0.68 & 0.62 & 0.42 & 9.32 & -0.32 \\

UGC05520 & 50.1 & 1.40 & 9.96 & 9.44 & 6.31 & 0.89 & 7.18 & 0.99 & 0.39 &
0.15 & 0.28 & 0.15 & 8.66 & -3.08 \\

UGC05976 & 22.6 & 0.57 & 8.89 & 8.29 & 5.80 & 0.91 & 6.55 & 1.03 & 0.27 &
0.12 & 0.35 & 0.23 & 8.28 & -3.03 \\

UGC06517 & 40.9 & 0.82 & 9.94 & 9.23 & 6.34 & 1.10 & 6.78 & 0.98 & 0.53 &
0.19 & 0.41 & 0.17 & 8.91 & -2.83 \\

UGC06616 & 22.2 & 0.52 & 9.27 & 8.25 & 5.78 & 1.01 & 6.22 & 0.97 & 0.40 &
0.17 & 0.41 & 0.27 & 8.33 & -2.61 \\

UGC06930 & 15.5 & 0.47 & 9.2 & 8.31 & 6.21 & 1.02 & 6.66 & 1.02 & 0.31 &
0.12 & 0.32 & 0.14 & 8.46 & -2.9 \\

UGC07012 & 49.4 & 0.82 & 9.97 & 9.17 & 6.91 & 1.07 & 6.36 & 1.06 & 0.42 &
0.14 & 0.31 & 0.15 & 8.85 & -2.41 \\

UGC08733 & 39.7 & 0.89 & 9.67 & 8.67 & 5.87 & 1.11 & 5.14 & 0.97 & 0.25 &
0.12 & 0.32 & 0.17 & 8.27 & -2.82 \\

UGC08781 & 113.0 & 3.75 & 11.28 & 10.95 & 11.93 & 0.60 & 10.08 & 0.86 & 0.89
& 0.59 & 0.74 & 0.36 & 9.31 & -0.24 \\

UGC08909 & 29.8 & 0.43 & 9.40 & 8.38 & 6.15 & 1.34 & 5.35 & 0.86 & 0.45 &
0.22 & 0.40 & 0.26 & 8.61 & -2.03 \\

UGC09067 & 116.0 & 1.81 & 11.1 & 10.49 & 11.61 & 0.51 & 8.85 & 0.86 & 0.65
& 0.29 & 0.50 & 0.25 & 9.47 & -0.41 \\

UGC09291 & 47.6 & 1.16 & 10.03 & 9.33 & 7.96 & 1.00 & 6.97 & 1.05 & 0.43 &
0.17 & 0.40 & 0.18 & 8.71 & -2.04 \\

UGC09476 & 52.3 & 1.69 & 10.47 & 9.82 & 8.50 & 1.00 & 7.30 & 1.09 & 0.54 &
0.18 & 0.51 & 0.21 & 8.87 & -1.51 \\

UGC09837 & 43.2 & 1.23 & 9.80 & 9.04 & 5.45 & 0.81 & 5.57 & 0.89 & 0.43 &
0.17 & 0.32 & 0.14 & 8.37 & -3.6 \\

UGC10796 & 48.0 & 0.99 & 9.67 & 8.91 & 5.47 & 1.11 & 4.54 & 0.95 & 0.37 &
0.13 & 0.31 & 0.17 & 8.42 & -3.39 \\

UGC11649 & 55.3 & 1.75 & 10.76 & 10.29 & 9.92 & 0.95 & 6.81 & 0.98 & 0.82 & 0.37 & 0.96 & 0.37 & 9.3 & -0.74 \\

UGC12224 & 48.7 & 1.29 & 10.28 & 9.45 & 7.82 & 1.03 & 7.47 & 1.04 & 0.57 &
0.23 & 0.48 & 0.22 & 8.73 & -1.97 \\

UGC12250 & 98.9 & 2.80 & 11.15 & 10.77 & 11.33 & 0.73 & 9.57 & 0.82 & 0.89 &
0.55 & 0.66 & 0.33 & 9.38 & -0.52 \\

UGC12767 & 71.2 & 2.57 & 11.43 & 10.9 & 10.70 & 0.70 & 8.36 & 0.93 & 0.65 &
0.34 & 0.53 & 0.26 & 9.58 & -0.63 \\

UGC12816 & 71.9 & 1.39 & 10.01 & 9.44 & 8.05 & 1.01 & 5.75 & 1.04 & 0.38 &
0.16 & 0.31 & 0.15 & 8.66 & -2.11 \\

UGC12864 & 63.6 & 1.35 & 10.23 & 9.70 & 8.32 & 1.01 & 6.08 & 1.03 & 0.48 &
0.19 & 0.40 & 0.17 & 8.94 & -2.05 \\

UGCA021 & 26.4 & 1.16 & 9.39 & 8.88 & 7.33 & 1.07 & 4.18 & 0.83 & 0.32 &
0.16 & 0.32 & 0.23 & 8.26 & -2.28 \\
\end{longtable}

\end{landscape}

}\end{appendix}
\end{document}